\documentstyle[prl,twocolumn,aps,epsf,eqsecnum,bbox]{revtex}
\begin{document}




\title{Transverse Momentum Dependence of the Landau-Pomeranchuk-Migdal Effect}
\author{Urs Achim Wiedemann and Miklos Gyulassy}

\address{Physics Department, Columbia University, New York,
NY 10027, USA}

\date{\today}
\maketitle

\begin{abstract}
We study the transverse momentum dependence of the 
Landau-Pomeranchuk-Migdal effect in QED, starting from the high energy 
expansion
of the solution of the Dirac equation in the presence of an external field. 
The angular integrated 
energy loss formula differs from an earlier expression of Zakharov by 
taking finite kinematical boundaries into account. In an expansion in 
powers of the opacity of the medium, we derive explicit expressions for 
the radiation cross section
associated with $N=1$, $2$ and $3$ scatterings. We verify
the Bethe-Heitler and the factorization limit, and we calculate
corrections to the factorization limit proportional to the square of 
the target size. A closed form expression valid to arbitrary orders
in the opacity is derived in the dipole approximation. The resulting
radiation spectrum is non-analytic in the coupling constant which
is traced back to the transverse momentum broadening of a hard parton 
undergoing multiple small angle Moli\`ere scattering. In extending
the results to QCD, we test a previously used dipole prescription
by comparing to direct pQCD results for $N=1$ and $2$. 
For $N=1$, the QCD dipole prescription reproduces
exactly the Bertsch-Gunion radiation spectrum. For $N=2$, we find a
sizeable correction which reduces to a multiplicative factor 17/8
at large separation.
\end{abstract} 


\section{Introduction}
\label{sec1}

In QED, 
the Landau-Pomeranchuk-Migdal (LPM) effect interpolates between the 
Bethe-Heitler and factorization limit for the radiation spectrum of a
charged particle undergoing multiple, $N$ say, scatterings in a medium.
If the separation between scattering centers is very large, then the
radiation off these centers reduces to a sum of radiation spectra
for $N$ small-angle scatterings - this is the Bethe-Heitler limit. 
In the opposite limit, when the
scattering centers sit too close together to be resolved by the
emitted photon, the observed radiation factorizes into a product
of a single scattering Bethe-Heitler spectrum for momentum transfer 
$q = \sum_{i=1}^N q_i$ and the elastic cross section for the
momentum transfer $q$ accumulated over $N$ small-angle scatterings. 

 As first noted by Landau and Pomeranchuk~\cite{LP53,Landau}, 
the relevant length scale
for the interpolation between Bethe-Heitler and factorization limit is
the coherence length (formation length) $l_f$, determined by the
longitudinal momentum transfer $q_l$,
\begin{equation}
  l_f = 1/q_l\, .
  \label{1.1}
\end{equation}
This characterizes the longitudinal scale on which the radiated particle
becomes distinguishable from its radiating parent. Scattering amplitudes
for the radiation off different scattering centers interfer 
destructively if their separation is less than $l_f$: the {\it coherent}
factorization limit is suppressed in the ultrarelativistic limit
with respect to the {\it incoherent} Bethe-Heitler limit. 

For a quantitative description of the LPM interference
effect, the relative phases of the different contributions to the 
$N$-fold scattering amplitude matter. These depend on the transverse
energies and thus require knowledge about the transverse motion of the 
radiating particle in the medium. Migdal~\cite{M56} was the first to develop
a dynamical description to this aim, employing a two-dimensional 
Fokker-Planck transport equation~\cite{M56,B58} for the hard parton. In the 
limit of an infinite medium, his well-known result shows a characteristic
$\propto \sqrt{\omega}$ low frequency suppression of $dN_\gamma/d\log\omega$
compared to the constant dependence in the Bethe-Heitler limit, i.e.,  
the coherent factorization limit for $\omega \to 0$ vanishes. This is, 
however, only a very special feature of Migdal's limiting case,
where the formation length goes to infinity for $\omega \to 0$, but
never exceeds the (infinite) extension $L$ of the medium. For a medium 
of finite size, the formation length does exceed the system size below 
some critical frequency $\omega_{\rm cr}$, and the Bethe-Heitler limit 
is finite. Migdals $\sqrt{\omega}$-dependence is hence not valid for 
$\omega < \omega_{\rm cr}$. Moreover, additional effects become important 
for the radiation spectrum at lower frequencies~\cite{K98}. Most notably,
this is the transition radiation and the Ter-Mikaelian effect~\cite{Ter53} 
of dielectric suppression. 

The renewed interest in the LPM-effect has at least two reasons: On the
one hand, fourty years after discovering the theoretical 
principles~\cite{LP53,M56}, the first precision measurements of the 
LPM-effect~\cite{SLAC1,SLAC3} (and the Ter-Mikaelian 
effect~\cite{SLAC3,SLAC2}) were made recently by the SLAC-146 
collaboration. On the other hand, with the advent of a new generation of 
relativistic heavy ion colliders at RHIC and LHC, the understanding of the
non-abelian analogue becomes important.

In the QED case, the experiment explored relatively thin targets 
with $L/l_f$ on the order 0.1 to 10, in which the transition between 
Bethe-Heitler and factorization limit occurs.
For a quantitative understanding, a realistic theory has 
to account for the finite extension of the target, the multiple 
elastic scatterings in the target, multiple photon emission, and possibly
additional complications like radiation off structured targets. 

There are at least three modern approaches, which can account in principle 
for the LPM-effect in realistic targets. They implement the eikonal 
approximation for the radiating hard particle in different ways: 
i) Blankenbecler and Drell~\cite{BD96,B97} (see also Baier and 
Katkov~\cite{BK98,BK99a,BK99b}) started from the solution 
of the Klein-Gordon equation for the charged particle 
in the presence of an external field. The solution was approximated to order
$1/E$ in a high energy expansion in which the eikonal path of the radiating 
particle is recovered. ii) Zakharov~\cite{Z96} proposed a light-cone path 
integral formulation which
describes in a coordinate-space representation the transverse momentum 
kicks on the eikonal path, mimicking the elastic scatterings
by an effective dipole cross section. Zakharov's work~\cite{Zak-QED,Z98} 
provides the
most accurate description of the measured data so far. As we shall see in 
section~\ref{sec2c}, his starting point is closely related to a high energy 
expansion of the Dirac equation in the presence of an external potential.
iii) A third approach is due to R. Baier et al. 
(BDMPS)~\cite{BDMPS1}, who started from the radiation amplitudes for $N$-fold 
scattering in the eikonal approximation. For QED, the consistency of 
their approach and the work of Blankenbecler and Drell can be checked
diagrammatically~\cite{DL98}. A discussion of the LPM effect also exists
for non-equilibrium conditions~\cite{KV96}.

In the QCD case, high $p_t \leq 10$ GeV
jets will be one of the new probes of the dense matter produced in 
relativistic heavy ion collisions.
Deviations from factorized perturbative QCD due to final state
rescattering are sensitive to the density and coupling in the
plasma and will affect key observables like the high momentum tails 
of single particle spectra. The corresponding study of QCD radiative
energy loss due to final state rescattering was initiated by Gyulassy 
and Wang~\cite{GW94,WGP95}. Recently, it has been extended most notably 
in a series of papers by BDMPS~\cite{BDMPS2,BDMPS3,BDMS,BDMS-Zak} using 
equal time perturbation theory. Also, Zakharov
has pointed out that his formalism can be adopted to QCD bremsstrahlung 
with a dipole prescription~\cite{Z96,Z98}, and 
the equivalence of Zakharov's formalism with the work of BDMPS was 
sketched~\cite{BDMS-Zak}. 

All these calculations of the LPM-effect in QCD, however, are (1) limited to
the cases of infinitely many or very few ($N < 3$) rescatterings of the parton,
and (2) they mainly focus on the angular integrated energy loss $dE/dx\, dL$ in
which the transverse momentum dependence of the radiation pattern is averaged
out. This energy loss, however, is not a good observable for QCD, because the
parton shower is not directly observable due to hadronization.
In addition, the QCD bremsstrahlung of hard jets must compete with 
the hard radiation associated with the jet production. 
To detect modifications of this hard vacuum bremsstrahlung spectrum due 
to final state rescattering requires knowledge about the angular
distribution of the spectrum. In QCD, the transverse momentum dependence
of the LPM-effect is thus indispensable for a quantitative understanding
of radiative energy loss.

Recently, Kopeliovich, Sch\"afer and Tarasov (KST)~\cite{KST98}
have used the Furry approximation of the Dirac equation in order to 
account for the transverse momentum dependence of the radiation spectrum 
emitted in a multiple scattering process. For the case of QCD bremsstrahlung
radiation, they translate their QED results into QCD via an
{\it a posteriori} dipole prescription (\ref{5.1}). This strategy is 
frequently used not only
for the calculation of the radiative energy loss of high $p_t$ 
partons~\cite{Z96,Z98,KST98}, but also for the description 
of nuclear shadowing in the target rest frame~\cite{KRT98,RTV98} 
and for related problems of diffractive dissociation of virtual 
photons~\cite{NZ94}.

As we show in what follows, an extension of the KST-formalism
has great potential for the calculation 
of radiative energy loss in realistic scenarios since (1) it provides
a smooth interpolation between the cases of infinitely many and
very few rescatterings and (2) it allows to compare QED-inspired
calculations of QCD radiative energy loss with perturbative QCD
results. The present work focuses on the general
formalism, its region of validity, and a qualitative discussion of
its generic features. It gives expressions which allow for the numerical
calculation of the radiation spectrum as a function of the medium
density and extension, but it leaves phenomenological applications to 
further publications.

Our work is organized as follows. In section~\ref{sec2}, we derive
the starting point of our discussion, the radiation spectrum (\ref{2.22}).
We discuss how a technical complication, the regularization of this 
spectrum, can be dealt with analytically, and we calculate the
corresponding integrated energy loss. Section~\ref{sec3} focuses on
limiting cases of the general
radiation formula (\ref{2.22}). We derive the radiation spectra for $N = 1$, 
$2$ and $3$ scatterings and we show that these reproduce the
Bethe-Heitler and the factorization limit. In 
section~\ref{sec4}, the dipole approximation of the
radiation spectrum (\ref{2.22}) is discussed. This
gives access to true in medium properties of the radiation spectrum 
which cannot be obtained from an expansion to finite order in the 
coupling constant. Finally, we discuss in 
section~\ref{sec5} how this method can be extended to QCD and how
it compares to results from perturbative QCD. Our main results are
summarized in the Conclusion.

\section{The LPM-effect in QED}
\label{sec2}

An expansion of the LPM-radiation cross section in orders of the 
coupling constant is essentially an expansion in the number of elastic
scatterings, since each elastic scattering Mott cross section is
proportional to $\alpha_{\rm em}^2$. In contrast, a high energy 
expansion of the solution of the Dirac equation in the 
presence of an external field takes to leading order
in $1/E$ an arbitrary number of elastic scatterings into account. 
In this section, we derive the corresponding high energy
limit of the QED radiation cross section for a hard electron, traversing 
a medium of longitudinal density $n(\xi)$. The physics contained in 
our main result (\ref{2.22}) will be discussed in the following
sections~\ref{sec3} - \ref{sec5}. Our approach in ~\ref{sec2a}
parallels to a large extent that of Kopeliovich, Sch\"afer 
and Tarasov (KST) in Ref.~\cite{KST98}. We present the derivation in 
full detail to introduce our notation and to discuss all the approximations 
involved in the calculation. In contrast to Ref.~\cite{KST98}, our
radiation cross section (\ref{2.22}) contains an $\epsilon$-regularization
whose treatment is discussed in subsection~\ref{sec2b}. Then we show
that the KST-formalism results in an angular integrated energy loss
formula from which modifications to Zakharov's formalism~\cite{Z96,Z98}
can be obtained.

\subsection{The KST-formalism: Differential cross section in the Furry 
approximation}
\label{sec2a}
We consider a relativistic electron undergoing multiple small-angle
scattering in a spatially extended medium, described by an
external potential $U({\bf x})$. The angular dependence of the
radiation spectrum for emitted photons carrying away a fraction 
$x$ of the incident electron energy $E_1$ is given by the
differential cross section
\begin{equation}
 \frac{d^5\sigma}{d({\rm ln}x)\,d{\bf p}_\perp\,d{\bf k}_\perp} =
 \frac{\alpha_{em}}{(2\pi)^4}\,
 \left|M_{fi}\right|^2\, ,
 \label{2.1}
\end{equation}
where ${\bf k}_\perp$ and ${\bf p}_\perp$ denote the transverse momenta 
of the photon and outgoing electron, respectively. The radiation 
amplitude for transversely polarized photons is given in terms of 
the ingoing and outgoing electron wavefunctions $\Psi^-$ and $\Psi^+$, 
\begin{equation}
  M_{fi} = \int
  d^4x\,{\Psi^-}^{\dagger}(x,p_2)\,
  \bbox{\alpha}\cdot \bbox{\epsilon}\,
  e^{-\epsilon\, |z|}\,
  e^{i\, k\cdot x}\,
  \Psi^+(x,p_1)\ .
  \label{2.2}
\end{equation}
Here, $\bbox{\alpha} = \gamma_0\, \bbox{\gamma}$, and $e^{-\epsilon\, |z|}$
is the adiabatic switching off of the interaction term at large distances.
This term plays an important role in what follows since the $\epsilon \to 0$
limit does not commute with the longitudinal $z$-integration. The wavefunctions
$\Psi^\pm$ solve the Dirac equation in the external potential $U({\bf x})$:
\begin{equation}
  \left[i\, {\partial\over \partial t} - U({\bf x}) - m\,\gamma_0 +
  i\,\bbox{\alpha}\cdot\bbox{\nabla}\right]\,
  \Psi(x,p_{1,2}) = 0\, .
  \label{2.3}  
\end{equation}
Following Kopeliovich et al.~\cite{KST98}, we rewrite the Dirac 
equation
\begin{eqnarray}
  &&\left[ - {\partial^2\over \partial^2 t} -m^2 + \Delta^2 -
       2\, i\, U({\bf x})\, {\partial\over \partial t}\right] \Psi(x,p_{1,2})
    \nonumber \\
  && = \left[ -i\, \bbox{\alpha}\cdot\left(\bbox{\nabla}U({\bf x})\right) 
       - U^2({\bf x})\right] \Psi(x,p_{1,2})\, .
  \label{2.4}
\end{eqnarray}
This allows for an expansion to order $1/E$, treating the right hand 
side as a small perturbation, solving the corresponding homogeneous 
problem and including in a first iteration the derivative term 
$\bbox{\nabla}U({\bf x})$. The result is the Furry 
approximation~\cite{KST98,Landau}:
 \begin{eqnarray}
  \Psi^+_F(x,p_1) &=& e^{-iE_1t + ip_1z}\,
   \hat D_1\,F^+({\bf x},{\bf p}_1)\,
   \frac{u({\bf p}_1)}{\sqrt{2\,E_1}} \, ,
   \label{2.5}\\
   \Psi^-_F(x,p_2) &=& e^{-iE_2t + ip_2z}\,
   \hat D_2\,F^-({\bf x},{\bf p}_2)\,
   \frac{u({\bf p}_2)}{\sqrt{2\,E_2}} \, ,
   \label{2.6}
 \end{eqnarray}
which is exact to order $O(U/E)$ and $O(1/E^2)$.
The building blocks of this solution are the 
differential operators $\hat D_i$ and the functions $F^{\pm}$
whose properties we explain now:

To determine the functions $F^{\pm}$, we neglect 
the second derivative $\partial^2/ \partial z^2$ in (\ref{2.4}). 
This is allowed since the longitudinal distances in a multiple 
scattering problem are much larger than the transverse ones. The
left hand side of (\ref{2.4}) reduces then to a two-dimensional 
Schr\"odinger equation, with corresponding retarded Green's function 
\begin{eqnarray}
 && \left[i\,\frac{{\it d}}{{\it d}z_2} +
 \frac{\Delta_\perp}{2\,p} - U(z_2,{\bf r}_2)\right]\,
 G(z_2,{\bf r}_2;z_1,{\bf r}_1|{\bf p}) \nonumber \\
 && \qquad \qquad =
 i\,\delta(z_2-z_1)\,\delta({\bf r}_2-{\bf r}_1)
 \label{2.7}\, ,
\end{eqnarray}
satisfying $G(z_2=z_1,{\bf r}_2;z_1,{\bf r}_1|{\bf p}) =
\delta({\bf r}_2-{\bf r}_1)$ and vanishing for $z_1>z_2$.
The functions $F^{\pm}$ in (\ref{2.5}), (\ref{2.6}) are solutions 
of this two-dimensional Schr\"odinger equation,
\begin{eqnarray}
 F^+({\bf x},{\bf p}_1) &=&
 \int d^2{\bf r}_1\,G(z,{\bf r};z_-,{\bf r}_1|{\bf p}_1)\,
 F^+({\bf r}_1,z_-,{\bf p}_1)\ ,\nonumber\\
 {F^-}^*({\bf x},{\bf p}_2) &=&
 \int d^2{\bf r}_2
 {F^-}^*({\bf r}_2,z_+,{\bf p}_2)\,
 G(z_+,{\bf r}_2;z,{\bf r}|{\bf p}_2)\, .
 \nonumber 
\end{eqnarray}
For very early and very late times, i.e. for far forward and far
backward longitudinal distances $z_+$ and $z_-$ respectively,
they satisfy the boundary conditions
\begin{eqnarray}
  F^+({\bf r}_1, z_-,{\bf p}_1) &=& 
  \exp\left\{i\,{\bf p}_{1\perp}\,{\bf r}_1
  - i\, {{\bf p}_{1\perp}^2\over 2\, p_1} z_-\right\}\, ,
  \label{2.8}\\
  F^-({\bf r}_2, z_+,{\bf p}_2) &=& 
  \exp\left\{i\,{\bf p}_{2\perp}\,{\bf r}_2
  - i\, {{\bf p}_{2\perp}^2\over 2\, p_2} z_+\right\}\, .
  \label{2.9}
\end{eqnarray}
This ensures that the wavefunctions $\Psi_F^{\pm}$ approximate
plane waves at asymptotic distances. The differential operators
$\hat{D}_i$ are obtained by including the derivative term 
$\bbox{\nabla}\, U({\bf x})$ of (\ref{2.4}) in a first 
iteration of this solution, 
 \begin{eqnarray}
 \hat D_i &=&
  1- i\,\frac{\bbox{\alpha}\cdot\bbox{\nabla}}{2\,E_i} 
   - \frac{\bbox{\alpha}\cdot({\bf p}_i-{\bf n}\,p_i)}{2\,E_i}\, ,
  \nonumber\\
  \bbox{n} &=& \frac{\bbox{k}}{k}\quad ;\quad
   z = \bbox{n}\cdot\bbox{x}\quad ;\quad
   p_i = |\bbox{p}_i|\, .
   \label{2.10}
\end{eqnarray}

\begin{figure}[t]\epsfxsize=7.5cm 
\centerline{\epsfbox{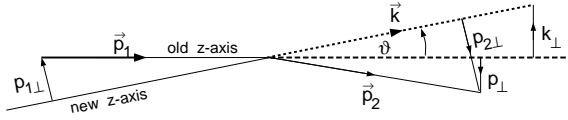}}
\vspace{0.5cm}
\caption{Choice of the coordinate system in which the new longitudinal
$z$-axis is taken along the radiated photon momentum, see Eqs.
(\protect\ref{2.11}), (\protect\ref{2.12}).}\label{fig4}
\end{figure}

We now explain how to obtain from the Furry approximation an explicit
expression of the differential radiation cross section (\ref{2.1}).
We work in the ultrarelativistic limit,
$E_1 \approx p_1$, $E_2 \approx p_2 = (1-x)\, p_1$, $\omega = x\, p_1$.
The longitudinal axis is redefined to be parallel to the emitted photon
$\bbox{k}$, see Figure~\ref{fig4}. 
The transverse momenta of the ingoing ($\bbox{p}_{1\perp}$)
and outgoing ($\bbox{p}_{2\perp}$) electron for small $x$ are 
therefore
\begin{eqnarray}
  \bbox{p}_{1\perp} &=& {-1\over x}\, {\bf k}_\perp  \, ,\qquad 
  \bbox{p}_{2\perp} = {\bf p}_\perp - {1-x\over x}\, {\bf k}_\perp\, ,
  \label{2.11}\\
  \bbox{q}_{\perp} &=& \bbox{p}_{2\perp} - \bbox{p}_{1\perp}
                    = \bbox{p}_{\perp} + \bbox{k}_{\perp}\, .
  \label{2.12} 
\end{eqnarray}
%
%
For the calculation of $M_{\rm fi}$, we use the following four steps:\\
1) We do the time-integral in (\ref{2.2}) which ensures 
energy conservation, $E_1 = E_2 + \omega$.\\
2) We determine the $z$-dependent phase factor $\bar{q}\, z$ for 
(\ref{2.2}) by observing that the $z$-dependent phases in (\ref{2.5}), 
(\ref{2.6}) come with the modulus of the spatial momentum 
$p_i = |\bbox{p}_i|$. Using energy-momentum conservation, 
one finds to leading order $1/E$ 
\begin{equation}
  \bar{q} = p_1 - p_2 - k = \frac{x\, m_e^2}{2\, (1-x)\, E_1}
  \equiv {1\over {l_f}}\, .
  \label{2.13}
\end{equation}
The inverse momentum transfer $1/\bar{q}$ is often refered to as
photon formation length $l_f$. However, contributions from transverse 
energies ${\bf k}_{\perp}^2 / 2\, \omega$ dominate the $z$-dependent
phase factor of ${\cal M}_{\rm fi}$, as will be seen below. Hence,
$l_f$ gives only a rough upper estimate for the longitudinal size over 
which interference effects suppress photon emission.\\
3) We absorb in (\ref{2.2}) the spinor structure of the wavefunctions 
$\Psi_F^+$, $\Psi_F^-$ in the interaction vertex
\begin{equation}
 \widehat\Gamma_r =
 \sqrt{1-x}\,u^*({\bf p}_2)\,\hat D^*_2\,
 \bbox{\alpha}\cdot \bbox{\epsilon}\,\hat D_1\,u({\bf p}_1)\, .
 \label{2.14}
\end{equation}
Depending on the electron spin of the ingoing
($\lambda_e = \pm \textstyle{1\over 2}$) and outgoing
($\lambda_{e'} = \pm \textstyle{1\over 2}$) electron
and the photon helicity ($\lambda_\gamma = \pm 1$), this vertex
takes the form
\begin{eqnarray}
  \widehat\Gamma_r\bigl(\lambda_e=\lambda_{e'}, \lambda_\gamma\bigr)
  &=& -i\,\lambda_\gamma\, 
      \left[ \lambda_\gamma\, (2-x) + 2\,\lambda_e\, x\right]
  \nonumber \\
   && \times \left( {\partial\over \partial r_x} 
             -i\, \lambda_\gamma\, {\partial\over \partial r_y}
             \right)\, ,
  \label{2.15} \\
  \widehat\Gamma_r\bigl(\lambda_e=-\lambda_{e'}, \lambda_\gamma\bigr)
  &=& 2\, m_e\, x\, \lambda_\gamma\, 
  \delta_{\lambda_\gamma\, , 2\lambda_e} \, ,
  \label{2.16}
\end{eqnarray}
where the spin-flip contribution (\ref{2.16}) is only non-vanishing for
$\lambda_\gamma = 2\, \lambda_e$. The differential operators in 
(\ref{2.14}) act to the right on the two different transverse
components of ${\bf r} = (r_x,r_y)$. In what follows, we are
only interested in the spin- and helicity-averaged 
combination $\widehat\Gamma_r\, \widehat\Gamma_{r'}^*$ which
takes the simple form
\begin{equation}
  \widehat\Gamma_r\, \widehat\Gamma_{r'}^* =
  \left[ 4-4x+2x^2\right] \frac{\partial}{\partial{\bf r}} \cdot
                          \frac{\partial}{\partial{\bf r'}}\, 
  + 2m_e^2x^2\, .
  \label{2.17}
\end{equation}
4) We split up the Green's functions
\begin{eqnarray}
  G(z_1,{\bf r}_1;z_2,{\bf r}_2) = \int d{\bf r}'
  G(z_1,{\bf r}_1;z',{\bf r}') G(z',{\bf r}';z_2,{\bf r}_2)
  \nonumber
\end{eqnarray}
at longitudinal distances $z'$ in such a way that contributions
from the amplitude (\ref{2.2}) and its complex conjugate part
can be paired, see eg. (\ref{2.18}) and Fig.~\ref{fig1}.

After these four steps, the radiation probability can be expressed as 
a Fourier transform
over pairs of Green's functions on whose transverse coordinates the 
interaction vertex (\ref{2.17}) is acting:
\begin{eqnarray}
 &&\langle|M_{fi}|^2\rangle = 2\, {\rm Re}\, 
 \int {\it d}{\bf r}_1\, {\it d}{\bf r}\, {\it d}\bbox{\rho}\, d{\bf r}_2\,
      {\it d}{\bf r}'_1\, {\it d}{\bf r}'\, 
      {\it d}\bbox{\rho}'\, {\it d}{\bf r}'_2
      \int dz 
      \nonumber \\
  &&\quad \times \int\limits_z^\infty dz'\, 
       e^{-i{\bf p}_{2\perp}\cdot({\bf r}_2-{\bf r}_2')
        +i{\bf p}_{1\perp}\cdot({\bf r}_1-{\bf r}_1')
        -i\bar{q}(z'-z)} e^{-\epsilon (|z|+|z'|)} \nonumber \\
  &&\quad \times \frac{1}{4\, p_1^2\, (1-x)^2}
                  \langle G(z_+,{\bf r}_2;z',\bbox{\rho}|p_2)\,
                  G^*(z_+,{\bf r}'_2;z',{\bf r}'|p_2)\rangle
   \nonumber \\
  &&\quad \times \widehat\Gamma_{-r}\, \widehat\Gamma_{r'}^*
      \langle G(z',\bbox{\rho};z,{\bf r}|p_2)\,
                  G^*(z',{\bf r}';z,\bbox{\rho}'|p_1)\rangle
   \nonumber \\
  &&\quad \times \langle G(z,{\bf r};z_-,{\bf r}_1|p_1)\,
                  G^*(z,\bbox{\rho}';z_-,{\bf r}'_1|p_1)\rangle\, .
  \label{2.18}
\end{eqnarray}
To turn this into an explicit expression, it is necessary to 
i) calculate the averages $\langle ...\rangle$ over the scattering 
centers in the medium and ii) evaluate the Green's functions $G$ 
which are given by path integrals
\begin{eqnarray}
  &&\qquad\qquad G(z',\bbox{r}';z,\bbox{r}\,|\,p) =  \nonumber \\
  &&\int {\cal D}\bbox{r}(\xi)\,
  \exp\left\{ \int\limits_{z}^{z'}{\it d}\xi\,
  \left[\frac{ip}{2} \dot{\bf r}^2(\xi) -
  i\,U\bigl({\bf r}(\xi),\xi\bigr)\right] \right\}
  \label{2.19}
\end{eqnarray}
with boundary conditions ${\bf r}(z)={\bf r}$, ${\bf r}(z')={\bf r}'$.
We show in Appendix~\ref{appa} that the in medium averages 
involved in (\ref{2.18}) result in a factor which depends on the cross section
$\sigma\left(\bbox{\rho}(\xi)\right)$ times the density
$n(\xi)$ of scattering centers as follows
 \begin{eqnarray}
   && 
   \Bigg\langle \exp\left\{ i\int\limits_{z}^{z'}{\it d}\xi\,  
    \left[ U\bigl({\bf r}(\xi),\xi\bigr) - U\bigl({\bf r}'(\xi),\xi\bigr) 
           \right]
    \right\}\Bigg\rangle \nonumber \\
   && \qquad = \exp\left\{ -  \int\limits_{z}^{z'}{\it d}\xi\,
       \Sigma\bigr(\xi,\left[\bbox{r}(\xi)-\bbox{r}'(\xi)\right]\bigl)  
                             \right\}\, , \label{2.20} \\
   &&\Sigma\bigl(\xi,\bbox{\rho}(\xi)\bigr) = \frac{1}{2}\, 
             n(\xi)\, 
             \sigma\bigl(\bbox{\rho}(\xi)\bigr)\, .
              \label{2.21}
 \end{eqnarray}
An explicit expression for $\sigma\bigl(\bbox{\rho}(\xi)\bigr)$,
supporting its interpretation as a Mott cross section, is also 
derived in Appendix~\ref{appa}. Most importantly, the resulting
expression $\Sigma\bigl(\xi,\bbox{\rho}(\xi)\bigr)$ depends only on the 
relative distance $\bbox{\rho}(\xi)$ between the paths ${\bf r}(\xi)$ 
and ${\bf r}'(\xi)$. As we explain in Appendix~\ref{appb}, this makes
it possible to carry out a 
large number of the integrals and path-integrals in (\ref{2.18})
analytically. One obtains (see Appendix~\ref{appb} for further details) 
\begin{eqnarray}
 &&\frac{d^5\sigma}{d({\rm ln}x)\,
   d{\bf p}_\perp\,d{\bf k}_\perp} =
 C_{\rm pre}\, {\rm Re} \int {\it d}\bbox{\rho}_2\,
 {\it d}\bbox{\rho}_1 \nonumber \\
 && \qquad 
    \times \int\limits_{z_-}^{z_+}\, dz \int\limits_{z}^{z_+}\, dz' 
    \exp\left\{-i\bar{q}(z'-z) -\epsilon(|z|+|z'|)  \right\}
    \nonumber \\
 && \qquad \times 
    \exp\left\{
    -\int\limits_{z_-}^z d\xi\,\Sigma(\xi,x\, \bbox{\rho}_1) 
    -\int\limits_{z'}^{z_+} d\xi\,\Sigma(\xi,x\, \bbox{\rho}_2)
    \right\}
    \nonumber \\
  && \qquad \times 
     \exp\left\{ -i\, x\, 
        \left({\bf p}_{\perp} - \frac{1-x}{x}\bbox{k}_{\perp}\right)
        \cdot \bbox{\rho}_2 
        -i{\bf k}_{\perp}\cdot \bbox{\rho}_1 
         \right\} 
     \nonumber \\ 
 && \qquad 
    \times 
    \left[ g_{\rm nf} {\partial\over \partial\bbox{\rho}_1}
                           \cdot {\partial\over \partial\bbox{\rho}_2}
        + g_{\rm sf} \right] 
        {\cal K}\bigl(z',\bbox{\rho}_2;z,\bbox{\rho}_1|\mu\bigr) \, .
    \label{2.22} 
\end{eqnarray}
This differs from the result in~\cite{KST98} by including the 
regularization $e^{-\epsilon(|z|+|z'|)}$.
The physical radiation spectrum is obtained by removing $\epsilon \to 0$
{\it after} doing the $z$ and $z'$-integrations and taking $z_-\to -\infty$
and $z_+\to \infty$.
To simplify the notation, we define 
\begin{equation}
  \mu \equiv E_1(1-x)x\, ,
  \label{2.22a}
\end{equation}
which plays the role of mass, and the factor
\begin{equation}
  C_{\rm pre} \equiv \frac{\alpha_{em}}
  {(2\pi)^4}\, {2\, x^2\over {E_1^2\, (1-x)^2}}\, ,
  \label{2.22b}
\end{equation}
which accounts for a combination of recurring kinematical prefactors.
The coefficients
of the spin-flip ($g_{\rm sf}$) and non-flip ($g_{\rm nf}$) contributions
of the interaction vertex (\ref{2.17}) read
\begin{equation}
  g_{\rm nf} = {{4-4x+2x^2}\over 4x^2}\, ,\qquad
  g_{\rm sf} = {m_e^2\, x^2\over 2}\, .
  \label{2.23}
\end{equation}
The remaining path integral ${\cal K}$ in (\ref{2.22}) is given by
\begin{eqnarray}
  &&{\cal K}\bigl(z',{\bf r}_c(z');z,{\bf r}_c(z)|\mu\bigr) 
  \nonumber \\
  && \quad =   \int {\cal D}{\bf r}_c\, 
  \exp\left\{i\, \int\limits_{z}^{z'}\, d\xi\,
  \left[{\mu\over 2}\dot{\bf r}_c^2 
  + i\, \Sigma\bigl(\xi,x\, {\bf r}_c\bigr)\right] \right\}\, .
  \label{2.24}
\end{eqnarray}
In this path integral, $\Sigma\bigl(\xi,x\, {\bf r}_c\bigr)$ plays
the role of an imaginary potential, while its physical interpretation is
up to a minus sign that of an elastic Mott cross section times the
density of scattering centers, cf. equation (\ref{a.12}). Depending
on the context, we shall hence refer to $\Sigma$ as a potential
or a cross section.

The following sections are devoted to a study of the differential
cross section (\ref{2.22}). We focus on analytically accessible
limiting cases which illustrate the physics contained in (\ref{2.22})
and we discuss approximations which allow for its numerical analysis.

\subsection{Removing the regularization $\epsilon \to 0$}
\label{sec2b}
The regularization prescription in (\ref{2.22}) cannot be
neglected since the $\epsilon \to 0$ limit does not commute with 
the $z$- and $z'$-integrations. This complicates practical 
applications of (\ref{2.22}): even if the Green's function ${\cal K}$ is
known explicitly in some approximation (see sections~\ref{sec3} 
and ~\ref{sec4} below), the expression 
(\ref{2.22}) is not suited for numerical calculations since 
one cannot control numerically the $\epsilon \to 0$-limit 
{\it after} carrying out the $z$-integrations.

To solve this problem, we determine the $\epsilon \to 0$
limit of the radiation spectrum (\ref{2.22}) analytically. We consider 
a medium of arbitrary but finite longitudinal extension which is positioned
along the longitudinal axis between 0 and the finite distance $L$.
The $z$- and $z'$-integrations of (\ref{2.22}) can be split into
six parts,
\begin{eqnarray}
  \int\limits_{z_-}^{z_+} \int\limits_{z}^{z_+}
  &=& \int\limits_{z_-}^{0} \int\limits_{z}^{0}
  + \int\limits_{z_-}^{0} \int\limits_{0}^{L}
  + \int\limits_{z_-}^{0} \int\limits_{L}^{z_+}
  \nonumber \\
  && + \int\limits_{0}^{L} \int\limits_{z}^{L}
  + \int\limits_{0}^{L} \int\limits_{L}^{z_+}
  + \int\limits_{L}^{z_+} \int\limits_{z}^{z_+}\, .
  \label{2.25}
\end{eqnarray}
To shorten the following calculations, we omit the spin-flip
contribution $g_{\rm sf}$ to the radiation spectrum. This contribution 
is negligible in the relativistic limit $m_e \ll E_1$ on which we focus 
in the following.
[If needed, the $g_{\rm sf}$-contributions to all our results can
be recovered without additional technical difficulties.]

The differential cross section receives six contributions which we 
label according to (\ref{2.25}) in an obvious way:
\begin{eqnarray}
 &&\frac{d^5\sigma}{d({\rm ln}x)\,d{\bf p}_\perp\,d{\bf k}_\perp} =
 C_{\rm pre}\, g_{\rm nf}\nonumber \\
 && \qquad \times
 \left( I_1 + I_2 + I_3 + I_4 + I_5 + I_6 \right)\, .
\label{2.26}
\end{eqnarray}
Since the medium does not
extend outside the interval $[0,L]$, we can replace the in-medium
propagator ${\cal K}\bigl(z',\bbox{\rho}_2;z,\bbox{\rho}_1|\mu\bigr)$
outside $[0,L]$ by the free propagator ${\cal K}_0$, given 
in (\ref{3.3}) below. E.g., for $z_2 > L > z_1$, we can write
\begin{eqnarray}
  {\cal K}\bigl(z_2,\bbox{\rho}_2;z_1,\bbox{\rho}_1|\mu\bigr)
  &=& \int d\bbox{r}\, 
    {\cal K}_0\bigl(z_2,\bbox{\rho}_2;L,\bbox{r}|\mu\bigr)
    \nonumber \\
  && \times {\cal K}\bigl(L,\bbox{r};z_1,\bbox{\rho}_1|\mu\bigr)\, .
  \label{2.27}
\end{eqnarray}
This allows to do all ``infinite'' integrals $\int_{z_-}^0$ and
$\int_L^{z_+}$ in (\ref{2.25}) analytically. One is left with the 
``finite'' integrals $\int_0^L$ for which the $\epsilon \to 0$ limit
commutes with the $z$-integration. 
Doing the $z$-integrations first, then taking $z_- \to -\infty$ and 
$z_+ \to +\infty$ respectively,
and finally adiabatically switching off the regularization, $\epsilon \to
0$, we find
\begin{eqnarray}
  I_1 &=& \left(x^2\, \bbox{p}_{1\perp}^2\right) \int {\it d}{\bf r}\,
      e^{ -\int_{0}^{L} \Sigma(\xi,x\, {\bf r}) 
      - ix\, {\bf q}_\perp\cdot {\bf r} }\, {\cal Z}_1\, ,
      \label{2.28} \\
  I_2 &=& - {\rm Re} \int\limits_0^L {\it d}z'\, e^{-i\, \bar{q}\, z'}
      \int {\it d}{\bf r}\, 
          e^{ -\int_{z'}^{L} \Sigma(\xi,x\, {\bf r}) 
      - ix\, {\bf p}_{2\perp}\cdot {\bf r} }\, {\cal Z}_2
      \nonumber \\
      && \times \int d\bar{\bf r}\, e^{ix\, {\bf p}_{1\perp}\cdot\bar{\bf r}}\,
      x{\bf p}_{1\perp}\cdot {\partial\over \partial \bbox{r}}
      {\cal K}(z',\bbox{r};0,\bar{\bf r}|\mu)\, ,
      \label{2.29} \\
  I_3 &=& (x{\bf p}_{1\perp})\cdot(x{\bf p}_{2\perp})\, {\rm Re} \int
      {\it d}{\bf r}_1\, {\it d}{\bf r}_2\,
      e^{-i\, x\, {\bf p}_{2\perp}\cdot {\bf r}_2} \,
      \nonumber \\
      && \times  e^{i\, x\, {\bf p}_{1\perp}\cdot {\bf r}_1}
                  {\cal K}(L,\bbox{r}_2;0,\bbox{r}_1|\mu)\, {\cal Z}_3\, 
                  e^{-i\, L\, \bar{q}},
      \label{2.30} \\
  I_4 &=& {\rm Re} \int\limits_{0}^{L} {\it d}z\, 
      \int\limits_{z}^{L} {\it d}z'\, \int {\it d}{\bf r}_1\,
      {\it d}{\bf r}_2\, e^{- i\, x\, {\bf p}_{2\perp}\cdot {\bf r}_2}
      e^{i\, x\, {\bf p}_{1\perp}\cdot {\bf r}_1}
      \nonumber \\
      && \times e^{ -\int_{0}^{z} \Sigma(\xi,x\, {\bf r}_1) 
         -\int_{z'}^{L} \Sigma(\xi,x\, {\bf r}_2)}
         e^{-i\, \bar{q}\, (z'-z)}
      \nonumber \\
      && \times   {\partial\over \partial {\bf r}_1}\cdot
         {\partial\over \partial {\bf r}_2}
         {\cal K}(z',\bbox{r}_2;z,\bbox{r}_1|\mu)\, ,
      \label{2.31} \\
  I_5 &=& {\rm Re} \int\limits_0^L {\it d}z\, e^{i\, \bar{q}\, (z-L)}
      \int {\it d}{\bf r}\, 
          e^{ -\int_{0}^{z} \Sigma(\xi,x\, {\bf r}) 
      + ix\, {\bf p}_{1\perp}\cdot {\bf r} }\, {\cal Z}_5
      \nonumber \\
      && \times \int d\bar{\bf r}\, 
      e^{-ix\, {\bf p}_{2\perp}\cdot\bar{\bf r}}\,
      x{\bf p}_{2\perp}\cdot {\partial\over \partial \bbox{r}}
      {\cal K}(L,\bar{\bf r};z,\bbox{r}|\mu)\, ,
      \label{2.32} \\
  I_6 &=& \left(x^2\, \bbox{p}_{2\perp}^2\right) \int {\it d}{\bf r}\,
      e^{ -\int_{0}^{L} \Sigma(\xi,x\, {\bf r}) 
      - ix\, {\bf q}_\perp\cdot {\bf r} }\, {\cal Z}_6\, .
      \label{2.33}
\end{eqnarray}
Here, the factors ${\cal Z}_i$ contain the explicit solutions of those
$z$-integrals which extend to $z_+$ or $z_-$ respectively. Using the
notational shorthands
\begin{eqnarray}
  Q_1 &=& {x^2 {\bf p}_{1\perp}^2\over 2\mu} + \bar{q} 
      = {x^2 \left( {\bf p}_{1\perp}^2 + m_{\rm e}^2\right)\over 2\mu} 
      \, , \label{2.34}\\
  Q_2 &=& {x^2 \left( {\bf p}_{2\perp}^2 + m_{\rm e}^2\right)\over 2\mu}\, ,
  \label{2.35}
\end{eqnarray}
we obtain
\begin{eqnarray}
  {\cal Z}_1 &=&  {\rm Re} \int\limits_{z_-}^{0}\, dz 
          \int\limits_{z}^{0}\, dz'\, 
          e^{-i\, Q_1\, (z'-z)
             - \epsilon(|z|+|z'|)}\,
           \nonumber \\
           &=&  {1\over 2\, Q_1^2}\, ,
 \label{2.36}
\end{eqnarray}
and analogously, 
\begin{eqnarray}
  {\cal Z}_2 &=& \frac{1}{Q_1}\, ,\qquad
  {\cal Z}_3 = - \frac{1}{Q_1\, Q_2}\, ,
  \label{2.37} \\
  {\cal Z}_5 &=& \frac{1}{Q_2}\, ,\qquad
  {\cal Z}_6 = {1\over 2\, Q_2^2}\, .
  \label{2.38}
\end{eqnarray}  
The differential radiation spectrum (\ref{2.26}) given by the 
equations (\ref{2.28}) - (\ref{2.38}) is suitable for a numerical
evaluation since the $\epsilon \to 0$-limit is taken. 
Previous numerical investigations~\cite{KST98} of 
(\ref{2.22}) were based on a non-regularized expression
which did not include the term $e^{-\epsilon(|z|+|z'|)}$ in the
integrand. The difference to the above results is found by
determining the above $z$-integrals without regularization:
\begin{eqnarray}
    {\cal Z}_1^{nr} &=&  {\rm Re} \int\limits_{z_-}^{0}\, dz 
          \int\limits_{z}^{0}\, dz'\, 
          e^{-i\, Q_1\, (z'-z)}\, 
          \nonumber \\
          &=&  2\, {\cal Z}_1\, \left( 1 - \cos \left\{Q_1\, z_-\right\}
                    \right)\, ,
           \label{2.39}\\
    {\cal Z}_2^{nr} &=& {\cal Z}_2\, \left( 1 - \exp\left\{i\,Q_1\, 
                    z_-\right\}\right)\, ,
          \label{2.40}\\
    {\cal Z}_3^{nr} &=& {\cal Z}_3\, \left( 1 - \exp\left\{i\,
                    Q_1\, z_-\right\}\right)
                    \nonumber \\
                    && \times
                    \left( 1 - \exp\left\{i\, Q_2\, 
                    (z_+-L)\right\}\right)\, , 
    \label{2.41}\\
    {\cal Z}_5^{nr} &=& {\cal Z}_5\, \left( 1 - \exp\left\{i\, Q_2\, 
                    (z_+-L)\right\}\right)\, , 
    \label{2.42}\\
    {\cal Z}_6^{nr} 
          &=&  2\, {\cal Z}_6\, \left( 1 - \cos \left\{Q_2\, (z_+-L)\right\}
                    \right)\, .
    \label{2.43}          
\end{eqnarray}  
The differential cross-section (\ref{2.22}) can be related 
to the unregularized one $\sigma^{(nr)}$:
\begin{eqnarray}
 &&\frac{d^5\sigma}{d({\rm ln}x)\,
   d{\bf p}_\perp\,d{\bf k}_\perp} =
   \frac{d^5\sigma^{(nr)}}{d({\rm ln}x)\,
   d{\bf p}_\perp\,d{\bf k}_\perp}
   \nonumber \\
 && \qquad - {C_{\rm pre}\over 2}\, \int {\it d}\bbox{\rho}\,
      \exp\left\{-\int\limits_{z_-}^{z^+} \Sigma(\xi,x\, \bbox{\rho}) 
      - ix\, \bbox{q}_\perp\cdot \bbox{\rho}\right\}
      \nonumber \\
 && \qquad \times \left[ { {\left(x^2\, \bbox{p}_{1\perp}^2\right) }
                  \over Q_1}
                 + { {\left(x^2\, \bbox{p}_{2\perp}^2\right) }
                  \over Q_2} \right]\, 
  +\, \Theta_{\rm corr}^{\rm osc}[z_-,z_+]\, .
  \label{2.44}
\end{eqnarray}
Here, the correction term $\Theta_{\rm corr}^{\rm osc}[z_-,z_+]$ 
summarizes those oscillating contributions which vanish if one averages
suitably in a numerical calculation over the boundaries $z_-$ and $z_+$ 
of the longitudinal integration. In addition, there are two terms which
do not depend on the boundaries $z_-$, $z_+$. They stem from the
terms ${\cal Z}_1^{nr}$ and ${\cal Z}_6^{nr}$ which differ by an
overall factor 2 from the regularized expressions ${\cal Z}_1$,
${\cal Z}_6$. These terms provide
the proper subtraction of the Weizs\"acker-Williams fields of the
ingoing and outgoing electron. As will become clear below, they are
non-negligible, and ensure e.g. the consistency of (\ref{2.22}) with 
the Bethe-Heitler spectrum in the appropriate limiting cases.

\subsection{Integrated energy loss cross section}
\label{sec2c}
The total energy loss cross section is calculated by integrating
(\ref{2.22}) over transverse momenta. To this aim,
we rewrite the differential cross section as a function
of the relative momentum transfer ${\bf q}_\perp = {\bf k}_\perp +
{\bf p}_\perp$ and the transverse momentum ${\bf k}_\perp$ of the 
emitted photon. Since there is no kinematical boundary on ${\bf q}_\perp$,
the integration over the transverse momentum of the electron is simple:
\begin{eqnarray}
 &&\frac{d^3\sigma}{d({\rm ln}x)\, d{\bf k}_\perp} =
 (2\, \pi)^2\, 
 {C_{\rm pre}\over x^2}\, {\rm Re} \int {\it d}\bbox{\rho}_1 
 \int\limits_{z_-}^{z_+}\, dz \int\limits_{z}^{z_+}\, dz' 
 e^{-i\bar{q}(z'-z)}
 \nonumber \\
 && \qquad \times 
    \exp\left\{ -\epsilon(|z|+|z'|) 
    -\int\limits_{z_-}^z d\xi\, \Sigma(\xi,x\, \bbox{\rho}_1) 
    -i{\bf k}_{\perp}\cdot \bbox{\rho}_1 
    \right\}
    \nonumber \\
 && \qquad 
    \times 
    \left[ g_{\rm nf} {\partial\over \partial\bbox{\rho}_1}
                           \cdot {\partial\over \partial\bbox{\rho}_2}
        + g_{\rm sf} \right] 
        {\cal K}\bigl(z',\bbox{\rho}_2=0;z,\bbox{\rho}_1|\mu\bigr) \, .
    \label{2.45} 
\end{eqnarray}
The integration over the photon transverse momentum is more complicated 
due to the finite kinematical boundaries for the photon, $|{\bf k}_\perp|
< \omega = x\, E_1$:
\begin{eqnarray}
 &&\frac{d\sigma}{d({\rm ln}x)} =
 \int\limits_0^{\omega} {\it d}{\bf k}_\perp\, 
 \frac{d^3\sigma}{d({\rm ln}x)\, d{\bf k}_\perp}\, . 
 \nonumber \\
 &&\quad = { {\alpha_{\rm em}}\over (2\pi)^2}
   { 2\, {\rm Re}\over E_1^2\, (1-x)^2}
 \int {\it d}\bbox{\rho}_1 
 {2\, \pi\, \omega\, J_1(\omega\,\rho_1)\over \rho_1}
 \nonumber \\
 && \qquad \times \int\limits_{z_-}^{z_+}\, dz \int\limits_{z}^{z_+}\, dz' 
 e^{-i\bar{q}(z'-z) -\epsilon(|z|+|z'|) }
 \nonumber \\
 && \qquad \times 
    \exp\left\{ -\int\limits_{z_-}^z \Sigma(\xi,x\, \bbox{\rho}_1) 
    \right\}
    \left[ g_{\rm nf} {\partial\over \partial\bbox{\rho}_1}
                           \cdot {\partial\over \partial\bbox{\rho}_2}
        + g_{\rm sf} \right] 
    \nonumber \\
 && \qquad \times 
    {\cal K}\bigl(z',\bbox{\rho}_2=0;z,\bbox{\rho}_1|\mu\bigr) \, .
 \label{2.46}
\end{eqnarray}
It is interesting to compare this energy loss formula to that 
derived by Zakharov~\cite{Z96,Z98}:
\begin{eqnarray}
 \frac{d\sigma^{({\rm zak})} }{d({\rm ln}x)} &=&
 \alpha_{\rm em}\, 
   { 2\, {\rm Re}\over E_1^2\, (1-x)^2}
 \int\limits_{z_-}^{z_+}\, dz \int\limits_{z}^{z_+}\, dz' 
 e^{-i\bar{q}(z'-z)}
 \nonumber \\ 
 && \times 
  \left[ g_{\rm nf} {\partial\over \partial\bbox{\rho}_1}
                           \cdot {\partial\over \partial\bbox{\rho}_2}
        + g_{\rm sf} \right] 
 \nonumber \\
 && \times \left[
       {\cal K}\bigl(z',0;z,0|\mu\bigr) 
       - {\cal K}_0\bigl(z',0;z,0|\mu\bigr) 
       \right]\, .
    \label{2.47} 
\end{eqnarray}
Here, ${\cal K}_0$ is the free, non-interacting path integral,
given explicitly in (\ref{3.3}) below. If we had ignored the kinematical 
boundary, extending the $k_\perp$-integration in (\ref{2.46}) up to 
infinity, we would have regained the expression of Zakharov except for
the term proportional to ${\cal K}_0$. 

We note that Zakharov's arguments
leading to (\ref{2.47}) are very different from our derivation of
(\ref{2.46}). The transverse momentum dependence of the radiation
enters in no intermediate step of his calculation~\cite{Z98}. Also, the
${\cal K}_0$-term in (\ref{2.47}) does not result from a derivation: 
it is rather subtracted {\it a posteriori} as a `renormalization 
prescription' in order to cancel a singularity in the integrand of 
(\ref{2.47}) for small $(z'-z)$. In this sense, our derivation differs
from Zakharov's result (\ref{2.47}) by: i) taking the finite phase space 
of the emitted photon properly into account, ii) containing a proper
$\epsilon$-regularization of the radiation amplitude and iii) not 
employing a subtraction of a singular contribution a posteriori.

\section{The low opacity expansion for thin targets}
\label{sec3}
Information about the target medium enters the radiation cross section
(\ref{2.22}) via the product $\Sigma(\xi,\bbox{\rho})$, which measures 
the elastic cross section times the density of scattering centers in 
the medium. Since $\Sigma(\xi,\bbox{\rho}) \propto \alpha_{\rm em}^2$
$n(\xi)$, an expansion of (\ref{2.22}) in powers of $\alpha_{\rm em}^2$ 
is an expansion in powers of the opacity ${\cal T}\, \sigma_{\rm eff}$,
where $ \sigma_{\rm eff}$ is the effective elastic single scattering
cross section and 
\begin{equation}
  {\cal T} = \int\limits_0^L\, d\xi\, n(\xi)\, .
  \label{3.1}
\end{equation}
${\cal T}\, \sigma_{\rm eff}$ measures the average number of scatterings 
for an electron traversing a medium of length $L$. Since the elastic cross
section in (\ref{2.22}) is not a geometrical quantity but depends on the
integration variables, our expansion of (\ref{2.22}) will be formally
in powers of $\alpha_{\rm em}^2\, {\cal T}$, and we shall refer by a
slight abuse of language to ${\cal T}$ as opacity. Of course, after
all integrals are done, each power of $\alpha_{\rm em}^2\, {\cal T}$
will be accompanied by a power of the elastic cross section.
The $N$-th term in this expansion of (\ref{2.22}) is
of order $\alpha_{\rm em}^{(2\, N + 1)}\, {\cal T}^N$ and corresponds 
to the multiple scattering off exactly $N$ external potentials.
An expansion in the opacity ${\cal T}$
is thus an expansion in the number of rescatterings. 

Below, we study this low opacity expansion of the radiation spectrum 
up to third order. This leads for equation (\ref{2.22}) to
a number of consistency checks which any radiation spectrum including
in medium effects should satisfy. Moreover, this provides the
basis for a discussion of the corresponding QCD radiation spectrum
in section~\ref{sec5}.

\subsection{Bethe-Heitler cross section as a low density limit}
\label{sec3a}
The derivation of the radiation spectrum (\ref{2.22}) relies on 
the approximation that the distribution of scattering centers
in the medium can be described by the average (\ref{2.20}). 
This uses explicitly that the size of the medium is much larger 
than the Debye radius of a single scattering
potential. A priori, it may hence seem unclear 
to what extent one can still recover from (\ref{2.22}) the correct 
radiation spectrum for a single scattering process where the extension
of the potential is the extension of the target. However, 
in the low opacity limit, when the distance between scattering 
centers is much larger than the photon formation length, multiple
scattering should not affect the radiation pattern. The radiation 
spectrum should converge to the single scattering Bethe-Heitler
cross section times the opacity factor ${\cal T}$. Deriving this
is an important consistency check for our formalism.

Our expansion of the integrand of the cross section (\ref{2.22}) 
in powers of $\Sigma(\xi,\bbox{\rho}) \propto \alpha_{\rm em}^2$
$n(\xi)$, uses the corresponding expansion 
of the path integral
\begin{eqnarray}
 &&{\cal K}(z',\bbox{\rho}_2;z,\bbox{\rho}_1) =
    {\cal K}_0(z',\bbox{\rho}_2;z,\bbox{\rho}_1)
 \nonumber \\
 && - \int\limits_{z}^{z'}\, {\it d}\bar{z} \int {\it d}\bbox{\rho}\,
 {\cal K}_0(z',\bbox{\rho}_2;\bar{z},\bbox{\rho})\,
   \Sigma\bigl(\bar{z},x\,\bbox{\rho}\bigr)\, 
   {\cal K}_0(\bar{z},\bbox{\rho};z,\bbox{\rho}_1) 
 \nonumber \\
 && + \int\limits_{z}^{z'} {\it d}\bar{z}_1\,
    \int\limits_{\bar{z}_1}^{z'} {\it d}\bar{z}_2\,
    \int {\it d}\bar{\bf r}_1\,{\it d}\bar{\bf r}_2
    {\cal K}_0(z',\bbox{\rho}_2;\bar{z}_2,\bar{\bf r}_2)\, 
    \Sigma\bigl(\bar{z}_2,x\,\bar{r}_2\bigr)
    \nonumber \\
 && \qquad \times 
    {\cal K}(\bar{z}_2,\bar{\bf r}_2;\bar{z}_1,\bar{\bf r}_1)\,
    \Sigma\bigl(\bar{z}_1,x\,\bar{r}_1\bigr)\,
    {\cal K}_0(\bar{z}_1,\bar{\bf r}_1;z,\bbox{\rho}_1)\, ,
 \label{3.2} 
\end{eqnarray}
where from now on, we suppress the explicit $\mu$-dependence in
${\cal K}$ and in the corresponding free Green's function ${\cal K}_0$, 
\begin{equation}
  {\cal K}_0(z',\bbox{\rho}_2;z,\bbox{\rho}_1)
  = \frac{\mu}{2\pi\, i\, (z'-z)}
    \exp\left\{ { {i\mu}
           \left(\bbox{\rho}_1 - \bbox{\rho}_2\right)^2
           \over {2\, (z'-z)} }
           \right\}\, .
  \label{3.3}
\end{equation}
From the two potentials in the third line of
(\ref{2.22}) and from the propagator ${\cal K}$, the radiation
spectrum (\ref{2.22}) receives three contributions to order
$O(\alpha_{\rm em}^3\, {\cal T})$:
\begin{eqnarray}
 &&\frac{d^5\sigma}{d({\rm ln}x)\,d{\bf p}_\perp\,d{\bf k}_\perp}
 \Big\vert_{O(\alpha_{\rm em}^3\, {\cal T})} =
 C_{\rm pre}\, g_{\rm nf}\, \left( I_1 + I_2 + I_3\right)\, ,
 \label{3.4}\\
  &&I_1 = - {\rm Re} \int\limits_{z_-}^{z_+}\, dz 
          \int\limits_{z}^{z_+}\, dz'\, 
          e^{-i\bar{q}(z'-z) -\epsilon(|z|+|z'|)}
          \int {\it d}\bbox{\rho}_2\, {\it d}\bbox{\rho}_1 
 \nonumber \\
 && \qquad \times 
    e^{ i\,x\,{\bf p_{2\perp}}\cdot{\rho_2}
                - i\,x\,{\bf p_{1\perp}}\cdot{\rho_1} } 
    \int\limits_{z_-}^z \Sigma(\xi,x\, \bbox{\rho}_1)\, 
    \kappa_1(\bbox{\rho}_1,\bbox{\rho}_2)\, ,
\label{3.5}\\
  &&I_2 = - {\rm Re} \int\limits_{z_-}^{z_+}\, dz 
          \int\limits_{z}^{z_+}\, dz'\, 
          e^{-i\bar{q}(z'-z) -\epsilon(|z|+|z'|)}
          \int {\it d}\bbox{\rho}_2\, {\it d}\bbox{\rho}_1 
 \nonumber \\
 && \qquad \times 
    e^{ i\,x\,{\bf p_{2\perp}}\cdot{\rho_2}
                - i\,x\,{\bf p_{1\perp}}\cdot{\rho_1} } 
    \int\limits_{z'}^{z_+} \Sigma(\xi,x\, \bbox{\rho}_1)\, 
    \kappa_1(\bbox{\rho}_1,\bbox{\rho}_2)\, ,
\label{3.6} \\
  &&I_3 = - {\rm Re} \int\limits_{z_-}^{z_+}\, dz 
          \int\limits_{z}^{z_+}\, dz'\, e^{-i\bar{q}(z'-z)}
          \int {\it d}\bbox{\rho}_2\, {\it d}\bbox{\rho}_1 
 \nonumber \\
 && \qquad\times 
    e^{ i\,x\,{\bf p_{2\perp}}\cdot{\rho_2}
                - i\,x\,{\bf p_{1\perp}}\cdot{\rho_1} } 
    \kappa_2(\bbox{\rho}_1,\bbox{\rho}_2)\, .
\label{3.7}
\end{eqnarray}
Here, $\kappa_1$ and $\kappa_2$ denote the derivatives of
the zeroth and first order contributions to the propagator (\ref{3.2}),
\begin{eqnarray}
  \kappa_1(\bbox{\rho}_1,\bbox{\rho}_2) &=&
  {\partial\over \partial\bbox{\rho}_1}
  \cdot {\partial\over \partial\bbox{\rho}_2}
  {\cal K}_0\bigl(z',{\bf \rho}_2;z,{\bf \rho}_1\bigr)\, .
\label{3.8}\\
  \kappa_2(\bbox{\rho}_1,\bbox{\rho}_2) &=&
  {\partial\over \partial\bbox{\rho}_1}
  \cdot {\partial\over \partial\bbox{\rho}_2}
  \int_{z}^{z'}\, {\it d}\bar{z} \int {\it d}\bbox{\rho}\,
 {\cal K}_0(z',\bbox{\rho}_2;\bar{z},\bbox{\rho})
 \nonumber \\
 && \qquad \qquad \quad \times
   \Sigma\bigl(\bar{z},x\,\bbox{\rho}\bigr)\, 
   {\cal K}_0(\bar{z},\bbox{\rho};z,\bbox{\rho}_1) \, .
 \label{3.9} 
\end{eqnarray}
We consider a medium of finite longitudinal extension $L$ and
homogeneous density $n(\xi) = n_0$. Then, all integrals in (\ref{3.5})
- (\ref{3.7}) can be done analytically. Taking the limit $\epsilon \to 0$
after the $z$-integrations, we arrive at
\begin{eqnarray}
  I_1 &=& - L\, n_0\,  \frac{x^2\, {\bf p}_{2\perp}^2}{2\, Q_2^2}
                  \frac{(2\pi)^2}{x^2}\, 
        \tilde{\Sigma}({\bf q}_\perp)\, ,
\label{3.10}\\
  I_2 &=& - L\, n_0\, \frac{x^2\, {\bf p}_{1\perp}^2}{2\, Q_1^2}
                  \frac{(2\pi)^2}{x^2}\, 
        \tilde{\Sigma}({\bf q}_\perp)\, ,
\label{3.11}\\
  I_3 &=& - L\, n_0\, 
        \frac{x^2\, {\bf p}_{2\perp}\cdot {\bf p}_{1\perp}}
               {Q_1\, Q_2}\, 
               \frac{(2\pi)^2}{x^2}\, \tilde{\Sigma}({\bf q}_\perp)\, .
\label{3.12}
\end{eqnarray} 
The result is particularly transparent in the relativistic limit, 
when all mass dependencies can be neglected and $\bar{q} \approx 0$.
For a medium characterized by a set of Yukawa
potentials with electric charge $Z\, e$ and Debye screening mass
$M$ (see Appendix~\ref{appa} for a derivation), 
\begin{equation}
  -\, \tilde{\Sigma}({\bf q}_\perp)\Bigg\vert_{{\bf q}_\perp \not= 0}
  = { 4\, Z^2\, \alpha_{\rm em}^2\over (M^2 + {\bf q}_\perp^2)^2}
  \label{3.13}
\end{equation}
denotes the elastic scattering cross section with ${\bf q}_\perp$
given by (\ref{2.12}). The three terms
(\ref{3.10})-(\ref{3.12}) then combine to
\begin{eqnarray}
 && \frac{d^5\sigma}{d({\rm ln}x)\,d{\bf p}_\perp\,d{\bf k}_\perp}
 \Bigg\vert_{O(\alpha_{\rm em}^3\, {\cal T})} \nonumber \\ 
 && \quad = \frac{Z^2\, \alpha_{\rm em}^3}{(2\, \pi)^2}\,
 \frac{16\, x^2}{(M^2 + {\bf q}_\perp^2)^2}\, 
 \frac{{\bf q}_\perp^2}{{\bf k}_\perp^2\,
              ({\bf k}_\perp - x\,{\bf q}_\perp)^2}\, {\cal T}\, .
 \label{3.14}
\end{eqnarray}
This is the high-energy limit of the Bethe-Heitler radiation
cross section for scattering off a Yukawa potential times 
the opacity factor ${\cal T} = L\, n_0$. The exact result
[see e.g. equation (5.150) of Ref.~\cite{IZ80} in the 
coordinate system Fig.~\ref{fig4} used here.] differs from 
the above approximation only by terms of order $1+O(q^2/E^2)$ and
$1+O(x^2)$. The main properties of the QED radiation spectrum are 
seen clearly from (\ref{3.14}). The spectrum vanishes for vanishing 
momentum transfer 
$q_\perp \to 0$, it shows the $x^2$-dependence characteristic for 
the QED radiation spectrum peaking at forward rapidity, and it shows 
the correct $1/{\bf k}_\perp^2$ $(\bbox{k}_\perp - x\bbox{q}_\perp)^2$ 
dependence.

We mention as an aside that the evaluation of the cross section
(\ref{3.4}) without regularization prescription leads
to terms $I_1$ and $I_2$ which are a factor 2 larger while the interference 
term remains unchanged. The $\epsilon$-regularization of (\ref{2.22})
is thus necessary to regain the Bethe-Heitler spectrum.

\subsection{Low N multiple scatterings}
\label{sec3b}
In complete analogy to the derivation of the Bethe-Heitler limit, one
can expand the cross section (\ref{2.22}) to second order.
All contributions to the radiation cross section depend 
now on the product of two elastic Mott cross sections $\tilde{\Sigma}$,
whose combined momentum transfers sum up to ${\bf q}_\perp$. For this
we introduce the shorthand
\begin{eqnarray}
  \int d{\cal V}^2(\bbox{q}_\perp) &\equiv& 
  \int d{\bf q}_{1\perp}\, d{\bf q}_{2\perp}\, 
  \tilde{\Sigma}({\bf q}_{1\perp})\, \tilde{\Sigma}({\bf q}_{2\perp})
  \nonumber \\
  && \times {(2\pi)^2\over x^2} \delta^{(2)}\left({\bf q}_\perp - 
                        {\bf q}_{1\perp} - {\bf q}_{2\perp}\right)\, .
  \label{3.15}
\end{eqnarray}
Also, we introduce shorthands for the transverse momenta and the 
corresponding transverse energies,
\begin{eqnarray}
  {\bf u}_1 &=& x\,{\bf p}_{1\perp}\, ,
  \qquad {\bf u}_2 = x\,{\bf p}_{2\perp}\, ,
  \nonumber \\
  {\bf u}_m &=& x\, \left( {\bf p}_{1\perp} + {\bf q}_{1\perp} \right)
             =  x\, \left( {\bf p}_{2\perp} - {\bf q}_{2\perp} \right)\, ,
  \nonumber \\
  Q_1 &=& { {\bf u}_1^2\over 2\mu}\, ,\qquad
  Q_m = { {\bf u}_m^2\over 2\mu}\, ,\qquad
  Q_2 = { {\bf u}_2^2\over 2\mu}\, .
  \nonumber 
\end{eqnarray}
The second order contribution of the radiation spectrum (\ref{2.22})
consists of six terms. After integrating out all transverse coordinates,
it takes the form
\begin{eqnarray}
 && \frac{d^5\sigma}{d({\rm ln}x)\,d{\bf p}_\perp\,d{\bf k}_\perp}
 \Bigg\vert_{O(\alpha_{\rm em}^5\, {\cal T}^2)} 
 = C_{\rm pre}\, g_{\rm nf}\, \int d{\cal V}^2(\bbox{q}_\perp)
 \nonumber \\
 && \qquad \times \left[ {1\over 2}\, {\bf u}_1^2 {\cal Z}_1^{(2)}  
            + {\bf u}_m^2 {\cal Z}_2^{(2)}
            + {1\over 2}\, {\bf u}_2^2 {\cal Z}_3^{(2)} 
            + {\bf u}_1\cdot{\bf u}_2 {\cal Z}_4^{(2)} \right.
          \nonumber \\
 && \qquad\qquad \left. 
            + {\bf u}_m\cdot{\bf u}_2 {\cal Z}_5^{(2)}
            + {\bf u}_m\cdot{\bf u}_1 {\cal Z}_6^{(2)}
            \right]\, n^2_0\, .
  \label{3.16}
\end{eqnarray}
The first three terms stem from the expansion of the exponential term
$\exp\left\{ -\int_{z_-}^z \Sigma - \int_{z'}^{z_+} \Sigma\right\}$ 
in (\ref{2.22}) to second order in $\Sigma$, 
the fourth term is from the second
order expansion of the Green's function ${\cal K}$, and the
remaining two terms are contributions from the first order in
${\cal K}$ times the first order of the exponential term. The
variables ${\cal Z}^{(2)}_i$ stand for the remaining longitudinal
integrals over phase factors. In general, they are involved, e.g.
\begin{eqnarray}
  {\cal Z}_4^{(2)} &=& {\rm Re} \int\limits_{z_-}^{z_+} dz\, 
  \int\limits_{z}^{z_+} dz'\, e^{-\epsilon(|z|+|z'|)}
  \int\limits_{z}^{z'} d\bar{z}_1\, {n(\bar{z}_1)\over n_0}
  \int\limits_{\bar{z}_1}^{z'} d\bar{z}_2\, {n(\bar{z}_2)\over n_0}\,
  \nonumber \\
  && \times
  e^{-iQ_2(z'-\bar{z}_2) - iQ_m(\bar{z}_2-\bar{z}_1)
     -iQ_1(\bar{z}_1-z)}\, .
  \label{3.17}
\end{eqnarray}
For a medium of homogeneous density $n_0$ and finite length $L$, they
reduce to very simple forms:
\begin{eqnarray}
  {\cal Z}_1^{(2)} &=& {L^2\over 2\, Q_1^2}\, ,
  \label{3.18} \\
  {\cal Z}_2^{(2)} &=& {{\cos(L\, Q_m) - 1 + 
                        {\textstyle{1\over 2}} L^2\, Q_m^2}\over Q_m^4}\, ,
  \label{3.19} \\
  {\cal Z}_3^{(2)} &=& {L^2\over 2\, Q_2^2}\, ,
  \label{3.20} \\
  {\cal Z}_4^{(2)} &=& {{\cos(L\, Q_m) - 1}\over Q_1\, Q_2\, Q_m^2}\, ,
  \label{3.21} \\
  {\cal Z}_5^{(2)} &=& - {{\cos(L\, Q_m) - 1 + 
                        {\textstyle{1\over 2}} L^2\, Q_m^2}
                        \over Q_2\, Q_m^3}\, ,
  \label{3.22} \\
  {\cal Z}_6^{(2)} &=& - {{\cos(L\, Q_m) - 1 + 
                        {\textstyle{1\over 2}} L^2\, Q_m^2}
                        \over Q_1\, Q_m^3}\, .
  \label{3.23}
\end{eqnarray}
From this, we read off simple limiting cases:
In the limit of a very thin target of fixed opacity, we can move the
two scattering centers so close together that the photon
cannot resolve them. The spectrum is then indistinguishable from 
a single scattering process
\begin{eqnarray}
 && \lim_{L\to 0} 
 \frac{d^5\sigma}{d({\rm ln}x)\,d{\bf p}_\perp\,d{\bf k}_\perp}
 \Bigg\vert_{O(\alpha_{\rm em}^5\, {\cal T}^2)}^{{\cal T}\, {\rm fixed}} 
  \nonumber \\
 && \quad = C_{\rm pre}\, g_{\rm nf}\, 
 { {\cal T}^2\over 4}
 \left( { {\bf u}_1\over Q_1} - {{\bf u}_2\over Q_2} \right)^2\, 
 \int d{\cal V}^2(\bbox{q}_\perp)\, .
  \label{3.24}
\end{eqnarray}
This is the coherent factorization limit that corresponds to one
single effective Bethe-Heitler radiation spectrum which depends
only on the initial and final momenta. 
The characterisitc momentum dependence of the Bethe-Heitler
radiation spectrum (\ref{3.14}) is now combined with the convolution of 
two Mott cross sections (\ref{3.15}). The first $L$-dependent
correction to this factorization limit is proportional to $L^2$
and takes the form
\begin{eqnarray}
 && C_{\rm pre}\, g_{\rm nf}\, 
 { {\cal T}^2\over 24}\, L^2
 \int d{\cal V}^2(\bbox{q}_\perp)\, Q_m^2
 \nonumber \\
 && \qquad \times
 \left[ 2\, {{\bf u}_m^2\over Q_m^2}
        + {{\bf u}_1\cdot {\bf u}_2\over Q_1\, Q_2}
        + {{\bf u}_m\cdot {\bf u}_2\over Q_m\, Q_2}
        + {{\bf u}_m\cdot {\bf u}_1\over Q_1\, Q_m} \right]\, ,
 \label{3.25}
\end{eqnarray}
which is mainly of formal interest since the 
${\bf q}_\perp$-integral diverges logarithmically. 

In the opposite limit, $L\to \infty$, we can study for fixed
opacity ${\cal T}= L\, n_0$ the case of two well-separated scattering
centers:
\begin{eqnarray}
 && \lim_{L\to \infty} 
 \frac{d^5\sigma}{d({\rm ln}x)\,d{\bf p}_\perp\,d{\bf k}_\perp}
 \Bigg\vert_{O(\alpha_{\rm em}^5\, {\cal T}^2)}^{{\cal T}\, {\rm fixed}} 
 = C_{\rm pre}\, g_{\rm nf}\, \int d{\cal V}^2(\bbox{q}_\perp)
 \nonumber \\
 && \qquad \times
 { {\cal T}^2\over 4} \left[
 \left( { {\bf u}_1\over Q_1} - {{\bf u}_m\over Q_m} \right)^2
 + \left( { {\bf u}_m\over Q_m} - {{\bf u}_2\over Q_2} \right)^2 \right]\, .
  \label{3.26}
\end{eqnarray}
This is the incoherent Bethe-Heitler limit in which the radiation 
spectrum is the sum of two independent Bethe-Heitler contributions 
associated with  
the scattering off the first and the second external potential with 
momentum transfers ${\bf q}_{1\perp}$ and ${\bf q}_{2\perp}$,
respectively. It provides the starting point for a comparison
of the corresponding QCD radiation spectrum with a full $N=2$
perturbative calculation in section~\ref{sec5}. 

The pattern emerging in this expansion to second order in the opacity
is fully confirmed in the case of $N = 3$ scattering centers. 
In Appendix~\ref{appc}, we derive the corresponding radiation cross
section and check the Bethe-Heitler and factorization limit.

\section{Dipole approximation for targets of arbitrary extension}
\label{sec4}

To calculate the radiation spectrum (\ref{2.22}) to all orders in
opacity, an approximation
scheme for the path integral ${\cal K}$ is needed. The low opacity
expansion, studied in section~\ref{sec3}, becomes rapidly more 
complicated for increasing number of scattering centers. It is 
only useful for the description of ultrathin media where very few 
elastic scatterings have to be taken into account. For the generic 
case of a hard projectile particle undergoing many small angle 
scatterings, the dipole approximation of ${\cal K}$ is 
standard~\cite{Z96,Z98,Zak-QED,KST98,NZ94,Z87}.

This dipole approximation is based on the observation that for small
transverse distances $\rho = |\bbox{\rho}|$, the cross section
$\sigma(\rho)$ of the Yukawa potential (\ref{a.11}) has a leading
quadratic dependence:~\cite{Z96}
\begin{eqnarray}
  \sigma(\bbox{\rho}) &\approx& C(\rho)\, \rho^2\, ,
  \label{4.1} \\
  C(\rho) &=& 4\,\pi\, (Z\,\alpha_{\rm em})^2\,
              \left[ \log\left(\frac{2}{M\,\rho}\right)
                     + \frac{1-2\,\gamma}{2} \right]\, ,
  \label{4.2}
\end{eqnarray}
where $\gamma = 0.577$ denotes Eulers constant. 
The main contribution to the radiation 
spectrum (\ref{2.22}) comes from small values of $\rho$, where 
$C({\rho}\, x)$ shows only a slow logarithmic dependence on 
$\rho$ and can be approximated by a constant, $C = C({\rho}_{\rm eff}\, x)$.
For a quadratic dependence $\sigma(\bbox{\rho}) = C\, \rho^2$, 
where this logarithmic dependence is weak enough to be neglected,
the path integral ${\cal K}$ in (\ref{2.24}) is that of 
a harmonic oscillator~\cite{Z96}
\begin{eqnarray}
  &&{\cal K}_{\rm osz}\bigl(z_2,{\bf r}_2;z_1,{\bf r}_1\bigr) 
     = {\mu\Omega\over 2\pi\, i\, \sin(\Omega\Delta z)}
  \nonumber \\
  && \times  \exp\left\{ 
     {i\mu\Omega 
     \left[ ({\bf r}_1^2 + {\bf r}_2^2)\cos(\Omega\Delta z)
                           -2\, {\bf r}_1\cdot{\bf r}_2\right]
    \over 2\, \sin(\Omega\Delta z)}
    \right\}
  \label{4.3}
\end{eqnarray}  
with the oscillator frequency
\begin{equation}
  \Omega = \frac{1-i}{\sqrt{2}}\, \sqrt{n\, C\, x^2\over \mu}\, .
  \label{4.4}
\end{equation}
In the following, we exploit the consequences of this approximation.

\subsection{The general expression: Non-analyticity in the coupling 
constant}
\label{sec4a}
In section~\ref{sec2b}, we have determined the six contributions
(\ref{2.28}) - (\ref{2.33}) to the radiation spectrum (\ref{2.22}).
In the dipole approximation, the transverse integrations in (\ref{2.28})
- (\ref{2.33}) reduce to Gaussian integrals and can be done analytically. 
Here we discuss the resulting expression 
\begin{eqnarray}
 &&\frac{d^5\sigma}{d({\rm ln}x)\,d{\bf p}_\perp\,d{\bf k}_\perp} =
 C_{\rm pre}\, g_{\rm nf}\,
 \sum_{j=1}^6\, I_j^{({\rm osc})}\, ,
\label{4.5}
\end{eqnarray}
obtained from (\ref{2.26}). For three of the 
six terms $I_j^{({\rm osc})}$, all integrals  can be done analytically:
\begin{eqnarray}
  I_1^{({\rm osc})} 
      &=& {1\over 2}\, {{\bf p}_{1\perp}^2\over Q_1^2}\,
          {2\,\pi\over n\, C\, L}\,
          \exp\left\{ -{ {\bf q}_{\perp}^2\over 2\, n\, C\, L}\right\}\, ,
          \label{4.6} \\
  I_6^{({\rm osc})} 
      &=& {1\over 2}\, {{\bf p}_{2\perp}^2\over Q_2^2}\,
          {2\,\pi\over n\, C\, L}\,
          \exp\left\{ -{ {\bf q}_{\perp}^2\over 2\, n\, C\, L}\right\}\, ,
          \label{4.7} \\
  I_3^{({\rm osc})}
      &=& {x^2\, {\bf p}_{1\perp}\cdot{\bf p}_{2\perp} \over Q_1\, Q_2}\,
          {\rm Re}\, e^{-i\, L\, \bar{q}}
          \nonumber \\
          && \times \exp\left\{ -  {i\, x^2\, \left({1-\cos(\Omega\, L)
              }\right)\over 2\, \mu\, \Omega\, \sin(\Omega\, L)}
            \left({\bf p}_{2\perp}^2 - {\bf p}_{1\perp}^2
            \right) \right\}
          \nonumber \\
          && \times {2\, i\, \pi\over \mu\, \Omega\, \sin(\Omega\, L)}\, 
          \exp\left\{ { i\, x^2\, {\bf q}_{\perp}^2\over 2\, \mu\, \Omega\,
            \sin(\Omega\, L) }\right\}\, .
          \label{4.8}
\end{eqnarray}
These terms show a characteristic dependence on the coupling constant:
in the dipole approximation, $C$ is constant and measures the size of the
elastic single scattering Mott cross section, see Eq. (\ref{4.2}). 
Thus, $C$ and $\Omega^2$ are proportional to $\alpha_{\rm em}^2$,
and the three terms $I_1$, $I_6$ and $I_3$ are non-analytic in the 
coupling constant. For small $\alpha_{\rm em}^2$, their leading 
dependence is $\alpha_{\rm em}^{-2}$ 
$\times \exp\left\{O(\alpha_{\rm em}^{-2})\right\}$. We show in 
the next subsection that these contributions 
combine to a radiation cross section with a leading dependence
$\alpha_{\rm em}$ $\times \exp\left\{O(\alpha_{\rm em}^{-2})\right\}$.
The factor
\begin{equation}
  \exp\left\{ -{ {\bf q}_{\perp}^2\over 2\, n\, C\, L}\right\}\, ,
  \label{4.9}
\end{equation}
responsible for the remaining non-analyticity of the radiation cross
section has an intuitive physical interpretation as Mol\`iere
${\bf q}_\perp$-broadening. This broadening is proportional
to the density of the medium $n_0$, the path length $L$ inside the medium
and the probability $C$ that a scattering center in the medium interacts
with the hard electron. As expected from a particle undergoing
random motion in the ${\bf q}_\perp$-plane, the accumulated
average ${\bf q}^2_\perp$ grows proportional to $L$. The corresponding 
contribution (\ref{4.9}) is seen explicitly in $I_1$ and $I_6$, but 
it can also be recovered from $I_3$ in an expansion to lowest order 
in $\alpha_{\rm em}^2$, when 
\begin{equation}
  i\, \mu\, \Omega\, \sin(\Omega\, L) = n\, C\, L \, x^2\, 
                                        + O(\alpha_{\rm em}^4)\, .
  \label{4.10}
\end{equation}
We now turn to the remaining three contributions of the radiation
cross section (\ref{4.5}), which are given in terms of integrals
over the longitudinal extension of the medium:
\begin{eqnarray}
  &&I_2^{({\rm osc})}  = {{\rm Re}\over Q_1} \int\limits_0^L {\it d}z\,
          e^{-i\,\bar{q}\,z}\,
          {\pi\, x^2\over C_2^2}\, \times
          \nonumber \\
          && \quad\, 
          \left[a_2^{(1)}\, {\bf p}_{1\perp}^2 + a_2^{(12)}\,  
          {\bf p}_{1\perp}\cdot{\bf p}_{2\perp} \right]\, \times
          \nonumber \\
          &&\quad 
          \exp\left\{{-x^2\over 4\, C_2}
           \left[ b_2^{(1)}\, {\bf p}_{1\perp}^2 
                  -2\,{\bf p}_{1\perp}\cdot {\bf p}_{2\perp}
                  + b_2^{(2)}\, {\bf p}_{2\perp}^2 \right]
                  \right\}\, ,
          \label{4.11} \\
  &&I_5^{({\rm osc})}  = {{\rm Re}\over Q_2} \int\limits_0^L {\it d}z\,
          e^{i\,\bar{q}\,(z-L)}\,
          {\pi\, x^2\over C_5^2}\,
          \nonumber \\
          && \quad\, 
          \left[a_5^{(2)}\, {\bf p}_{2\perp}^2 + a_5^{(12)}\,  
          {\bf p}_{1\perp}\cdot{\bf p}_{2\perp} \right]\, \times
          \nonumber \\
          &&\quad
          \exp\left\{{-x^2\over 4\, C_5}
           \left[ b_5^{(1)}\, {\bf p}_{1\perp}^2 
                  -2\,{\bf p}_{1\perp}\cdot {\bf p}_{2\perp}
                  + b_5^{(2)}\, {\bf p}_{2\perp}^2 \right]
                  \right\}\, ,
          \label{4.12}\\
  &&I_4^{({\rm osc})}  = -\, {\rm Re} \int\limits_0^L {\it d}z\,
          \int\limits_z^L {\it d}z'\, 
          e^{-i\,\bar{q}\,(z'-z)}
          {\pi\, x^2\over C_4^3}\, \times 
          \nonumber \\
          &&\quad \left[ a_4^{(0)} + a_4^{(1)}\, {\bf p}_{1\perp}^2 
                  + a_4^{(12)}\, {\bf p}_{1\perp}\cdot{\bf p}_{2\perp}
                  + a_4^{(2)}\, {\bf p}_{2\perp}^2\right]\, \times
          \nonumber \\
          &&\quad 
          \exp\left\{{-x^2\over 4\, C_4}
           \left[b_4^{(1)}\, {\bf p}_{1\perp}^2 
                   - 2\, {\bf p}_{1\perp}\cdot {\bf p}_{2\perp} 
                   + b_4^{(2)}\, {\bf p}_{2\perp}^2\right]
                   \right\}\, .
         \label{4.13}
\end{eqnarray}
To make the structure of these terms transparent, 
we have introduced several shorthands $a_i$, $b_i$ and $C_i$. 
The factors $b_i$ and $C_i$ show simple limiting behaviour:
\begin{eqnarray}
  b_i &=& 1\, + O(\alpha_{\rm em}^2)\, ,
  \label{4.15}\\
  4\, C_i &=& 2\, n\, C\, L\, x^2\, + O(\alpha_{\rm em}^4)\, .
  \label{4.14} 
\end{eqnarray}
This allows us to recover the Moli\`ere ${\bf q}_\perp$-broadening term
(\ref{4.9}) to leading order in the coupling constant in all six 
contributions $I_j^{({\rm osc})}$ of the radiation spectrum (\ref{4.5}).

The explicit form of the functions $C_i$ are:
\begin{eqnarray}
  C_2 &=& {i\, \mu \over 2}\, \Omega\,
          \left\{ \Omega\, (L-z)\, \cos[\Omega\, z] + \sin[\Omega\, z]
                 \right\}\, ,
           \label{4.16} \\
  C_5 &=& {i\, \mu\over 2}\, \Omega\,
          \left\{ \Omega\, z\, \cos[\Omega(L-z)] + \sin[\Omega(L-z)]
                 \right\}\, ,
           \label{4.17} \\
  C_4 &=& {i\, \mu\over 2}\,\Omega\, \left\{
         -\Omega^2\,z\,(L-z')\, \sin[\Omega(z'-z)] + \sin[\Omega(z'-z)]
         \right.
         \nonumber \\
         && \left. + 
         \Omega\,(L-z'+z)\, \cos[\Omega(z'-z)] \right\}\, ,
         \label{4.18}
\end{eqnarray}
from which the limit (\ref{4.14}) is recovered easily. The functions
$b_i$ are given by
\begin{eqnarray}
  b_2^{(1)} &=& \cos[\Omega\, z] -  \Omega\, (L-z)\, \sin[\Omega\, z]\, ,
            \label{4.19} \\
  b_2^{(2)} &=& \cos[\Omega\, z]\, ,
            \label{4.20} \\
  b_5^{(1)} &=& \cos[\Omega\, (L-z)]\, ,
            \label{4.21} \\
  b_5^{(2)} &=& \cos[\Omega\, (L-z)] -  \Omega\, z\, 
                \sin[\Omega\, (L-z)]\, ,
            \label{4.22} \\
  b_4^{(1)} &=& \cos[\Omega\, (z'-z)] - \Omega\, (L-z')\, \sin[\Omega\, (z'-z)]
          \, , \label{4.23}\\
  b_4^{(2)} &=& \cos[\Omega\, (z'-z)] - \Omega\, z\, \sin[\Omega\, (z'-z)]
          \, . \label{4.24}
\end{eqnarray}
from which the limiting form (\ref{4.15}) is also clear.
Finally, the prefactors of the integrands in (\ref{4.11})-(\ref{4.13})
are:
\begin{eqnarray}
  a_2^{(1)} &=& {\mu\over 2}\, \Omega^2 (L-z)\, ,\qquad
  a_2^{(12)} = {\mu\over 2}\, \Omega\, \sin[\Omega\, z]\, ,
  \label{4.25} \\
  a_5^{(2)} &=& {\mu\over 2}\, \Omega^2 z\, ,\qquad
  a_5^{(12)} = {\mu\over 2}\, \Omega\, \sin[\Omega\, (L-z)]\, ,
  \label{4.26} \\
  a_4^{(0)} &=& \mu^2\, \Omega^4\, C_4\, {1\over x^2}\, z(L-z')\, ,
  \label{4.27} \\
  a_4^{(1)} &=& {\mu^2\, \Omega^4\over 4}\, 
                \left( (L-z')^2\, c + {{L-z'}\over \Omega}\, s\right)\, ,
  \label{4.28} \\
  a_4^{(2)} &=& {\mu^2\, \Omega^4\over 4}\, 
                \left( z^2\, c + {z\over \Omega}\, s\right)\, ,
  \label{4.29} \\
  a_4^{(12)} &=& {\mu^2\, \Omega^4\over 4}\, 
                \left\{ z(L-z') + \left(z\, c + {s\over \Omega}\right)
                       \right.
             \nonumber \\
             && \qquad \qquad \qquad \times \left. 
                        \left((L-z')\, c + {s\over \Omega}\right)
                        \right\}\, ,
  \label{4.30}
\end{eqnarray}
where we have used $c \equiv \cos[\Omega(z'-z)]$ and $s \equiv
\sin[\Omega(z'-z)]$. The expressions given here allow for a 
numerical calculation of the radiation spectrum,
once the extension $L$ of the medium, its density $n_0$, and the 
measure $C$ of the elastic cross section of a single scatterer
are given. Based on these expressions, we discuss in the next
subsections
i) an analytically accessible limiting case of the
spectrum (\ref{4.5}) and ii) a Bethe-Heitler limit which allows
to determine the constant $C$ phenomenologically.

\subsection{Moli\`ere limit: Radiation in the Gaussian small angle 
multiple scattering regime}
\label{sec4b}
We now turn to a limiting case of the radiation spectrum
(\ref{2.22}) in which a characteristic in-medium effect can be
isolated in a simple analytical expression. This limit focusses
on the kinematical region
\begin{eqnarray}
   && q_\perp \ll k_\perp \ll k \ll E\, ,
   \label{4.31} \\
   && Q\, L = {k^2_\perp\over 2\, k}\, L \gg 1\, ,
   \label{4.32} \\
   && \bar{q}\, L \gg 1\, ,
   \label{4.33} \\
   && |\Omega\, L| \ll 1\, .
   \label{4.34}
\end{eqnarray}
The first condition (\ref{4.31}) is satisfied in the high energy limit
when $x$ is small and the photon is radiated under a small angle with
respect to the beam. It is characteristic for relativistic kinematics 
that the transverse momentum freed by multiple interactions can be 
significantly larger than the total transverse momentum $q_\perp$ 
transfered by the medium, i.e., ${\bf p}_\perp \approx - {\bf k}_\perp$ 
and $q_\perp = |{\bf p}_\perp + {\bf k}_\perp| \ll k_\perp$. According
to (\ref{4.32}) and (\ref{4.33}), the length of the medium has to
exceed one formation length significantly, while the last condition
$|\Omega\, L| \ll 1$ can be realized e.g. by choosing a target of
sufficiently low density. We note that (\ref{4.33}) is by far the
most stringent condition.

 As a consequences of (\ref{4.32}) and (\ref{4.33}), the
terms $\cos(Q\, L)$, $\sin(Q\, L)$ are rapidly oscillating as
function of $L$. If we assume a target with varying extension
$L$, e.g. due to an unpolished surface, then contributions to the 
radiation spectrum (\ref{4.5}) proportional to these oscillating terms 
will be averaged out by the experiment. An
expansion of the terms in (\ref{4.5}) in powers of the coupling 
constant leads then to
\begin{eqnarray}
  I_1^{({\rm osc})} 
      &=& {1\over 2}\, {{\bf p}_{*\perp}^2\over Q^2}\,
          {2\,\pi\over n\, C\, L}\,
          \exp\left\{ -{ {\bf q}_{\perp}^2\over 2\, n\, C\, L}\right\}\, ,
          \label{4.35} \\
  I_2^{({\rm osc})} 
      &=& \left[ {-\, 2\,\pi\over n\, C\, L}{{\bf p}_{*\perp}^2\over Q^2}
                  - {4\pi\over Q^2}{1\over L^2\, Q^2} 
                 \right]
          e^{ -{ {\bf q}_{\perp}^2\over 2\, n\, C\, L}}\, ,
          \label{4.36} \\
  I_3^{({\rm osc})}
      &=& 0
      \label{4.37} \\
  I_4^{({\rm osc})} 
      &=& \left[ {2\,\pi\over n\, C\, L}\,
                     {{\bf p}_{*\perp}^2\over Q^2} 
                     + {2\pi\over Q^2} 
                     + {{4\, \pi}\over Q^2}{1\over L^2\, Q^2} \right]
          e^{ -{ {\bf q}_{\perp}^2\over 2\, n\, C\, L}}\, ,
          \label{4.38} \\
  I_5^{({\rm osc})} 
      &=& \left[ {-\, 2\,\pi\over n\, C\, L}{{\bf p}_{*\perp}^2\over Q^2}
                  - {4\pi\over Q^2}{1\over L^2\, Q^2} 
                 \right]
          e^{ -{ {\bf q}_{\perp}^2\over 2\, n\, C\, L}}\, ,
          \label{4.39} \\
  I_6^{({\rm osc})} 
      &=& {1\over 2}\, {{\bf p}_{*\perp}^2\over Q^2}\,
          {2\,\pi\over n\, C\, L}\,
          \exp\left\{ -{ {\bf q}_{\perp}^2\over 2\, n\, C\, L}\right\}\, .
          \label{4.40} 
\end{eqnarray}
Here, we have used (\ref{4.32}) to neglect the ${\bf q}_\perp$-dependent 
terms except for the leading $O(1/\alpha_{\rm em}^2)$ Moli\`ere
factor (\ref{4.9}), and we have approximated
${\bf p}_{1\perp} \approx {\bf p}_{2\perp} \approx {\bf p}_{*\perp}$.
Corrections to these terms are of order $O({q_\perp^2\over k_\perp^2})$
and $O(1/(L^2\,Q^2)^2)$. Neglecting $O(1/(L^2\,Q^2))$-contributions,
the radiation spectrum (\ref{4.5}) takes the simple form 
\begin{eqnarray}
 \frac{d^5\sigma}{d({\rm ln}x)\,d{\bf p}_\perp\,d{\bf k}_\perp} &=&
 C_{\rm pre}\, g_{\rm nf}\,
 \sum_{j=1}^6\, I_j^{({\rm osc})}
 \nonumber \\
 &\approx& {\alpha_{\rm em}\over \pi^3}\, {x^2\over {\bf k}_\perp^4}\,
 \exp\left\{ -{ {\bf q}_{\perp}^2\over 2\, n\, C\, L}\right\}\, .
\label{4.41}
\end{eqnarray}
This Moli\`ere limit is the main result of the present subsection.
In accordance with
the Bethe-Heitler single scattering cross section (\ref{3.14}), we 
still find the characteristic $x^2$ rapidity dependence, as well as
the $1/{{\bf k}_\perp^4}$ fall off of the spectrum in the kinematical
regime $q_\perp \ll k_\perp$. In contrast to the single scattering
Bethe-Heitler cross section, however, the spectrum now peaks at
vanishing total momentum transfer $q_\perp = 0$ rather than to
vanish for $q_\perp \to 0$. The reason is that the radiation from many
small angle scatterings adds up to a finite contribution, while the
sum $\sum_i^N {\bf q}_{i\perp}$ of many small random momentum transfers 
${\bf q}_{i\perp}$  peaks at zero.

Also the dependence of (\ref{4.41}) on the coupling constant is
easy to understand. The cross section is proportional to $\alpha_{\rm em}$,
since one photon is radiated off. The broadening of the ${\bf q}_\perp$-
distribution is determined by the probability that the hard electron
undergoes an interaction. This probability is given by the elastic
single scattering Mott cross section, and hence the exponent shows
a characteristic $\alpha_{\rm em}^{-2}$-dependence. 

The radiation cross section (\ref{4.5}) obtained in the dipole approximation,
is clearly more complex than the limiting case (\ref{4.41}). The latter,
however, illustrates most clearly that generic in medium effects are
retained in the dipole approximation which cannot be obtained from a
calculation to fixed order in the coupling constant.
%
\subsection{Fixing the dipole parameter $C$ in the Bethe-Heitler limit:
validity of the dipole approximation}
\label{sec4c}

In the dipole approximation, the expansion of the radiation cross
section (\ref{2.22}) in powers of the coupling constant does not
converge. The spectra for $N=1,2,3$ scatterings derived in 
section~\ref{sec3}, are proportional 
to the $N$-th power of the elastic scattering cross section
$\tilde{\Sigma}({\bf q}_\perp)$ which is essentially the Fourier transform
of $\sigma(\bbox{\rho})$. While the Fourier transform of the analytic
expression (\ref{a.11}) for $\sigma(\bbox{\rho})$ is well-defined,
the Fourier transform of its dipole approximation 
$\sigma(\bbox{\rho}) = C\, \bbox{\rho}^2$ diverges.

This failure of the dipole approximation to allow for an expansion
of (\ref{2.22}) in powers of the opacity ${\cal T}$ does not affect
its validity for the calculation of medium effects as given e.g.
in the last subsections. This can be seen e.g. from the factor
$\exp\left\{ -\int \Sigma(\xi,x\,\bbox{\rho})\right\}$
in (\ref{2.22}). Its Fourier transform is well-defined and it is 
well approximated in the dipole approximation 
$\sigma(\bbox{\rho}) = C\, \bbox{\rho}^2$ by a Gaussian of appropriate 
width. It is only the expansion of this factor in $\alpha_{\rm em}$
whose Fourier transform does not receive the main contributions
from small values around $\rho=0$. This is the reason why the
dipole approximation of $\sigma(\bbox{\rho})$ for small values of 
$\bbox{\rho}$ cannot be combined without difficulties with an 
expansion in the coupling constant.

For the ${\bf q}_\perp$-integrated spectrum (\ref{2.45}), this
technical difficulty does not exist. The reason is that in the
integrand of (\ref{2.45}), the first order term $\Sigma(\xi,x\,\bbox{\rho}_1)$
comes multiplied by ${\cal K}(z',0;z,\bbox{\rho_1}|\mu)$ which is
to lowest order in the coupling constant a Gaussian in $\bbox{\rho}_1$.
This ensures that the main contribution to the integral (\ref{2.45})
comes from small values of $\bbox{\rho}$ where the dipole approximation
is valid. The corresponding Bethe-Heitler spectrum in which the
dipole approximation is employed {\it after} integrating out the
${\bf q}_\perp$-dependence, reads:
\begin{eqnarray}
 \frac{d^3\sigma}{d({\rm ln}x)\,d{\bf k}_\perp}
 \Bigg\vert_{O(\alpha_{\rm em}^3\, {\cal T})}^{\rm dipole}  
 &=&  {\alpha_{\rm em}\over 2\, \pi^2}\, x^2\, C\, 
   \left\{ 4-4x+2x^2\right\}
   \nonumber \\
   && \times {{{\bf k}^4_\perp + (m_e\, x)^4}
     \over { \left({{\bf k}^2_\perp + (m_e\, x)^2}\right)^4}}\, 
     {\cal T}\, .
 \label{4.42}
\end{eqnarray}
In the calculation of (\ref{4.42}) we have not neglected the
$\bar{q}$-dependence of the radiation spectrum (\ref{2.22}). 
The consequence is the apprearance of the very small term $m_e\, x$  
which regulates the $1\over {\bf k}_\perp^4$ singularity. This allows
to calculate the ${\bf k}_\perp$-integrated Bethe-Heitler energy loss 
formula in the dipole approximation 
\begin{eqnarray}
 &&\frac{d\sigma}{d({\rm ln}x)}
 \Bigg\vert_{O(\alpha_{\rm em}^3\, n)}^{\rm dipole}  
 =  {\alpha_{\rm em}\, C\over 3\, \pi\, m_e^2}\,  
   \left\{ 4-4x+2x^2\right\}
   \nonumber \\
   &&\qquad  \times 
   \left( 1 - m_e^2x^2 { {3\omega^4 + 3\omega^2\, m_e^2x^2 + m_e^4x^4}
     \over { 2\, (\omega^2 + m_e^2x^2)^3}}\right)\, .
 \label{4.43}
\end{eqnarray}
We note that except for the small correction from the kinematical
boundary $\omega = E_1\, x$ of the ${\bf k}_\perp$-integral, 
expression (\ref{4.43}) is consistent with the Bethe-Heitler term 
derived by Zakharov in the dipole approximation~\cite{Z98}. 
Equations (\ref{4.42}) or (\ref{4.43}) allow to determine the only
free parameter $C$ from a comparison with well-tabulated Bethe-Heitler
scattering cross section~\cite{T74}. Once, $C$ is fixed,
the radiation spectrum (\ref{4.5}) in the dipole approximation 
provides thus a parameter free prediction of the measured in-medium
energy loss. Zakharov has pursued this strategy to fix $C$ in his 
${\bf q}_\perp$- and ${\bf k}_\perp$- integrated spectrum (\ref{2.47}),
and this has lead to a very successful description of the SLAC-146
data on radiative energy loss~\cite{Z98}.

\section{The dipole prescription for QCD}
\label{sec5}

\begin{figure}[h]\epsfxsize=7.5cm 
\centerline{\epsfbox{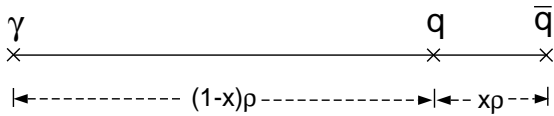}}
\vspace{0.5cm}
\caption{Separation of dipoles in the transverse plane: the 
$\bar{q}$-$\gamma$ fluctuation of transverse size $\rho$
gives rise to a QED radiation spectrum determined by the $\bar{q}$-$q$
dipole of size $x\, \rho$. In QCD, where the radiated gluon is
charged, all three dipole contributions should be considered.
More details are given in (\protect\ref{5.1}) and the following
text.}\label{fig2}
\end{figure}

To calculate the integrated QCD radiative energy loss of a hard 
coloured parton, Zakharov~\cite{Z98} has used a very simple substitution
in his energy loss formula (\ref{2.47}). This QCD dipole prescription 
consists in replacing the dipole cross section $\sigma(x\, \bbox{\rho})$ 
in the calculation of the QED radiation
spectrum by a combination of three dipole cross sections~\cite{NZ94}:
\begin{eqnarray}
  \sigma_{\rm QED}(\bbox{\rho},x) &=& \sigma(x\, \bbox{\rho})
  \quad \longrightarrow
  \nonumber \\
  \sigma_{\rm QCD}(\bbox{\rho},x) &=&
  {9\over 8}\, \left\{ \bar{\sigma}(\bbox{\rho}) +  
                    \bar{\sigma}((1-x)\, \bbox{\rho})\right\}
  - {1\over 8}\, \bar{\sigma}(x\, \bbox{\rho})\, .
  \label{5.1}
\end{eqnarray}
Here, we have chosen $\bar{\sigma}$ proportional to the elastic
$q$-$q$ Mott cross section, cf. subsection~\ref{sec5a}.
The heuristic argument for the prescription (\ref{5.1}) starts from the 
representation of the projectile quark in the light cone frame as a 
superposition of the bare quark and higher Fock states,
\begin{equation}
  \vert {\rm projectile}\rangle = \vert q\rangle + \vert q\,\gamma \rangle
                                 + \dots\, .
  \label{5.2}
\end{equation}
If all Fock components interact with the external potentials with the
same amplitude, then the coherence between these amplitudes is not
disturbed, and no bremsstrahlung is generated. The radiation amplitude
depends hence on the difference between the elastic scattering 
amplitudes of different fluctuations. The dipole cross section
$\sigma(x\,\bbox{\rho})$ contains information about this difference
since it arises from averaging in (\ref{2.20}) the (part of the) amplitude 
$M_{\rm fi}$ for the "ingoing" $\vert q\rangle$ with the (part of the)
complex conjugated amplitude $M_{\rm fi}^*$ for the "outgoing" higher 
Fock state $\vert q\,\gamma \rangle$. In the transverse plane, the
separation of the $\vert q\,\gamma \rangle$ fluctuation from the
ingoing $q$ can be estimated via the uncertainty relation:
$x_{\perp\, \gamma} \propto \textstyle{k_\perp\over x\, E_1}
\textstyle{1\over \Delta E} \propto (1-x)/k_\perp$ and
$x_{\perp\, q} \propto \textstyle{k_\perp\over (1-x)\, E_1}
\textstyle{1\over \Delta E} \propto x/k_\perp$. The transverse
distance $\bbox{\rho}$ is linked to the transverse momentum in (\ref{2.22})
by a Fourier transform and we may think of $\bbox{\rho}$
as the transverse size of the $\bar{q}$-$\gamma$ fluctuation.
With the transverse center of mass of this fluctuation at the
position of the ingoing quark $\vert q\rangle$, see Figure~\ref{fig2}, 
the transverse distance between the charged components $q$ and $\bar{q}$
is then $x\,\bbox{\rho}$, and it is the dipole of this size which determines
the radiation spectrum. For related arguments, see also 
Ref.~\cite{BHQ97}. 

%
In QCD, the emitted gluon is charged too, and
aside from the $q$-$\bar{q}$ dipole of size $x\,\bbox{\rho}$, there
is a $g$-$\bar{q}$ dipole of size $\bbox{\rho}$ and a $g$-$q$ dipole
of size $(1-x)\,\bbox{\rho}$. The prefactors introduced for these dipoles 
in the prescription (\ref{5.1}) stem from interpolating between 
the limits
\begin{eqnarray}
  \lim_{x\to 0} \sigma_{\rm QCD}(\bbox{\rho},x) &=& 
                {9\over 4}\, \bar{\sigma}(\bbox{\rho})\, ,
                \label{5.3}\\
  \lim_{x\to 1} \sigma_{\rm QCD}(\bbox{\rho},x) &=& 
                \bar{\sigma}(\bbox{\rho})\, .
                \label{5.4}
\end{eqnarray}
In the limit $x\to 0$, the $q\, \bar{q}$ pair is indistinguishable from
a pointlike color-octet charge and the ratio $9/4$ of the octet
to triplet couplings arises. In the opposite limit $x\to 1$, the
gluon-quark pair at vanishing separation is indistinguishable from 
a pointlike quark.

The question to what extent this intuitive physical picture 
leads to the correct QCD radiation cross section was addressed
recently in Ref.~\cite{BDMS-Zak}. There it is argued that Zakharov's 
integrated 
energy loss formula (\ref{2.47}) coincides for QCD with the result of 
their calculation based on time-ordered perturbation theory. For a 
discussion of the terms neglected in this approach, see 
Ref.~\cite{GLV99}. Here, we contribute to this discussion by
comparing the transverse momentum dependence resulting from
the QCD dipole prescription (\ref{5.1}) to the transverse momentum dependent
radiation cross sections for $N=1$ and $N=2$ scattering centers,
calculated~\cite{GLV99} in time-ordered perturbation theory.

\subsection{The Bertsch-Gunion spectrum for $N=1$ in the QCD
dipole prescription}
\label{sec5a}

In the present subsection, we test the QCD dipole prescription 
(\ref{5.1}) for the case of $N=1$ scattering. We start from
the $N=1$ QED Bethe-Heitler spectrum (\ref{3.14}) in the form
\begin{eqnarray}
 &&\frac{d^5\sigma^{\rm BH-QED}}{d({\rm ln}x)\,d{\bf q}_\perp\,
   d{\bf k}_\perp}
 = \frac{4\, \alpha_{\rm em}}{(2\, \pi)^2}\,
 \frac{x^2\, {\bf q}_\perp^2}{{\bf k}_\perp^2\,
              ({\bf k}_\perp - x\,{\bf q}_\perp)^2}\, {\cal T}
 \nonumber \\
 && \qquad \qquad 
 \times {-1\over 2}\int {d(x\bbox{\rho})\over (2\pi)^2}\, 
 \sigma_{\rm QED}(x\,\bbox{\rho})\, 
 \exp\left\{i\, x\, {\bf q}_\perp\cdot \bbox{\rho}\right\}\, .
 \label{5.5}
\end{eqnarray}
To specify the absolute size of $\bar{\sigma}$ in (\ref{5.1}),
we relate the elastic QED $e^-$-$e^-$ cross section to the 
corresponding QCD one by
\begin{eqnarray}
  {-1\over 2}\int {d{\bf r}\over (2\pi)^2}\, 
 \sigma({\bf r})\, e^{i\, {\bf q}_\perp\cdot {\bf r}}
 &=& {{4\, Z^2\, \alpha_{\rm em}^2}\over (M^2 + {\bf q}_\perp^2)^2}
 \nonumber \\
 \longrightarrow
  {-1\over 2}\int {d{\bf r}\over (2\pi)^2}\, 
 \bar{\sigma}({\bf r})\, e^{i\, {\bf q}_\perp\cdot {\bf r}}
 &=& C_i\, {{4\, \alpha_{\rm s}^2}\over (M^2 + {\bf q}_\perp^2)^2}\, ,
 \label{5.6}
\end{eqnarray}
where $C_i$ denotes the colour factor, $2\, C_i = $ $\textstyle{4\over 9}$,
$1$, $\textstyle{9\over 4}$, for $q$-$q$, $q$-$g$, $g$-$g$. Also, to
change to the QCD-case, we replace the coupling to the emitted
photon in (\ref{5.5}) by $\alpha_{\rm em} \to \alpha_{\rm s}\, C_A$,
where the Casimir $C_A$ of the adjoint representation accounts for 
the emitted gluon. With this input, we follow the dipole prescription
and substitute (\ref{5.1}) in the Bethe-Heitler cross section (\ref{5.5}).
After rescaling ${\bf q}_\perp$ $\to \textstyle{1\over x}\, {\bf q}_\perp$
on both sides of the equation, we find (note that $\bar{\sigma}(\rho) =
\bar{\sigma}((1-x)\, \rho) + O(x)$)
\begin{eqnarray}
 &&\frac{d^5\sigma^{\rm BG-QCD}}{d({\rm ln}x)\,d{\bf q}_\perp\,
   d{\bf k}_\perp}
 = \frac{C_A\, \alpha_{\rm s}}{\pi^2}\,
 \frac{ {\bf q}_\perp^2}{{\bf k}_\perp^2\,
              ({\bf k}_\perp - {\bf q}_\perp)^2}\, {\cal T}
 \nonumber \\
 && \qquad \times {8\over 9}\, \frac{\alpha_{\rm s}^2}
 {(M^2 + {\bf q}_\perp^2)^2}\, + O(x)\, .
 \label{5.7}
\end{eqnarray}
This is the Bertsch-Gunion radiation spectrum~\cite{BG82} times the elastic
Mott cross section for a Debye-screened coloured potential. For
$N=1$, the QCD dipole prescription turns to leading order  $O(1/E)$
the Bethe-Heitler QED radiation spectrum {\it exactly} into the 
Bertsch-Gunion QCD radiation spectrum. This success
does not depend on the particular combination of colour dipoles 
introduced in (\ref{5.1}). It is rather due to the scaling property
of the dipole size in (\ref{5.5}): replacing in the argument of $\sigma$ the 
transverse extension $x\, \bbox{\rho} \to \bbox{\rho}$ modifies
both: (i) the $x^2$ rapidity distribution of the QED spectrum into
the flat rapidity distribution of the QCD spectrum and (ii) the
characteristic $x\, {\bf q}_\perp$-dependence of the QED spectrum
into the characterisitc ${\bf q}_\perp$-dependence of the QCD
radiation spectrum.

\subsection{Corrections to the QCD dipole prescription for $N=2$}
\label{sec5b}

The interference terms of QCD multiple scattering bremsstrahlung
radiation are more complicated than their QED analogue. Especially,
since the gluon is charged, it can rescatter on additional potentials,
and these cascading contributions do not factorize from the gluon
production amplitudes. This effect does not exist for the QED 
bremsstrahlung: there, the only in medium modification of the produced
photons is the dielectric suppression which is clearly a final state
effect which factorizes from the production amplitude. 

\begin{figure}[t]\epsfxsize=7.5cm 
\centerline{\epsfbox{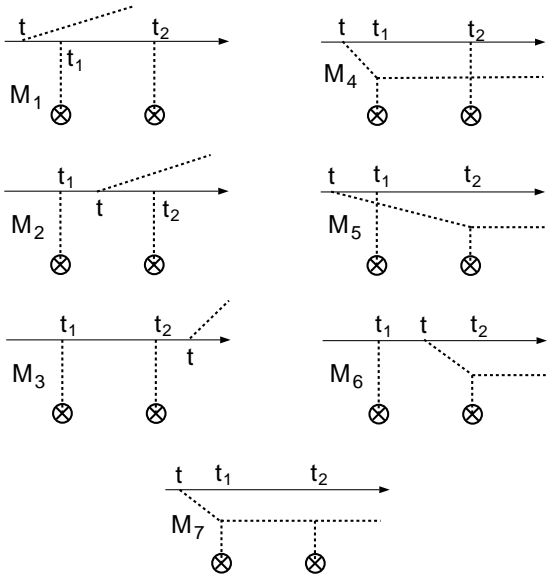}}
\vspace{0.5cm}
\caption{The seven contributions to the $N=2$ QCD radiation spectrum
in time-ordered perturbation theory. In the large distance limit,
the Bertsch-Gunion and gluon cascading contributions can be identified
with particular combinations of these diagrams, see equations 
(\protect\ref{5.14}) - (\protect\ref{5.16}) and the text.
}\label{fig3}
\end{figure}

This qualitatively different structure of QCD rescattering processes
makes it interesting to study the accuracy of the QCD dipole prescription
for $N=2$ scatterings. To this aim, we compare the $N=2$ radiation
spectrum (\ref{3.15}) to the result of the full $N=2$ QCD radiation
spectrum calculated in time ordered perturbation theory in Ref.~\cite{GLV99}:
\begin{eqnarray}
  &&\frac{d^5\sigma^{\rm full\, QCD}_{N=2}}{d({\rm ln}x)\,d{\bf q}_\perp\,
   d{\bf k}_\perp}
 = {C_A\, \alpha_{\rm s}\over \pi^2}\, 
 \int d{\cal V}^2_{\rm QCD}({\bf q}_\perp)\, 
 \nonumber \\
 &&\times \left\{ \vec{B}^2_1 + \vec{B}^2_2 +
   R\, \vec{B}^2_{2(12)} 
  - {C_A\over 2\, C_F} \left(
   \vec{B}_1\cdot \vec{B}_2\, \cos(\omega_1\, \Delta t) \right. \right.
  \nonumber \\ 
  && \qquad \left. \left. -2\, \vec{B}_2\cdot \vec{B}_{2(12)} 
     \cos(\omega_2\, \Delta t) \right. \right.
  \nonumber \\
  && \qquad \left. \left. 
     + \vec{B}_1\cdot \vec{B}_{2(12)} \cos(\omega_{20})\, \Delta t)
   \right)
    \right\}\, ,
    \label{5.8}
\end{eqnarray}
where $\omega_0 = \textstyle{{\bf k}_\perp^2\over 2\, \omega}$,
$\omega_2 = \textstyle{({\bf k}_\perp-{\bf q}_{2\perp})^2\over 2\, \omega}$,
$\omega_{20} = \omega_2 - \omega_0$. Equation (\ref{5.8}) is obtained
by squaring the amplitude in Eq. (2.20) of Ref.~\cite{BDMPS2} which
corresponds to the seven diagrams in Fig.~\ref{fig3}. We have 
used the following shorthands:
\begin{eqnarray}
  &&\int d{\cal V}^2_{\rm QCD}({\bf q}_\perp)
   =  \int d{\bf q}_{1\perp}\, d{\bf q}_{2\perp}\,
      { \textstyle{8\over 9}\, \alpha_{\rm s}^2\over 
        (M^2 + {\bf q}_{1\perp}^2)^2}\, 
      \nonumber  \\
  && \qquad \times { \textstyle{8\over 9}\, \alpha_{\rm s}^2\over 
        (M^2 + {\bf q}_{2\perp}^2)^2}\, 
      \delta^{(2)}({\bf q}_\perp- {\bf q}_{1\perp} - {\bf q}_{2\perp})\, ,
     \label{5.9}\\
  &&\vec{B}_i = { {\bf k}_\perp\over {\bf k}_\perp^2}
                - { {\bf k}_\perp - {\bf q}_{i\perp}\over 
                    ({\bf k}_\perp - {\bf q}_{i\perp})^2}\, ,
                  \label{5.10} \\
  &&\vec{B}_{i(12)} = { {\bf k}_\perp - {\bf q}_{i\perp}\over 
                    ({\bf k}_\perp - {\bf q}_{i\perp})^2}
                  - { {\bf k}_\perp - {\bf q}_{1\perp} - {\bf q}_{2\perp} 
                    \over ({\bf k}_\perp - {\bf q}_{1\perp}
                           - {\bf q}_{2\perp})^2}\, .
                  \label{5.11}
\end{eqnarray}
The $N=2$ radiation spectrum (\ref{5.8}) is obtained for the case of
two scattering centers placed at {\it fixed} longitudinal positions
$z_1 = t_1$, $z_2 = t_2$ with $\Delta t = t_2 - t_1$. In contrast,
the $N=2$ calculation presented in section~\ref{sec3b} considers
two scattering centers at {\it arbitrary} longitudinal positions 
distributed according to a homogeneous density $n_0$ within a
longitudinal extension $L$. In fact, the interference terms in
(\ref{5.8}) depend in a different way on the kinematical variables
than the interference terms in (\ref{3.16}). Moreover, we cannot compare the
$\Delta t \to 0$ limit of both expressions, since (\ref{5.8}) is 
obtained in time ordered perturbation theory where $t_1 < t_2$ is
used to neglect diagrammatic contributions. This leaves us for a
direct comparison of both calculations with only the limiting case 
\begin{eqnarray}
  &&\lim_{\Delta t \to \infty}
  \frac{d^5\sigma^{\rm full\, QCD}_{N=2}}{d({\rm ln}x)\,d{\bf q}_\perp\,
   d{\bf k}_\perp}
 ={C_A\, \alpha_s\over \pi^2}\, \int d{\cal V}^2_{\rm QCD}({\bf q}_\perp)
 \nonumber \\
 && \qquad \times \left\{ \vec{B}^2_1 + \vec{B}^2_2 +
   {C_A\over C_F}\, \vec{B}^2_{2(12)}\right\}\, ,
  \label{5.12}
\end{eqnarray}
which corresponds to two scattering centers at arbitrary large 
relative distance. One checks that the QCD dipole prescription 
leads for the $N=2$ rescattering result (\ref{3.26}) in the 
$L\to\infty$ limit  
\begin{eqnarray}
  &&\lim_{L \to \infty}
  \frac{d^5\sigma^{\rm dipole\, QCD}_{N=2}}{d({\rm ln}x)\,d{\bf q}_\perp\,
   d{\bf k}_\perp}
 ={C_A\, \alpha_s\over \pi^2}\, \int d{\cal V}^2_{\rm QCD}({\bf q}_\perp)
 \nonumber \\
 && \qquad \times \left\{ \vec{B}^2_1 + \vec{B}^2_{1(12)}\right\}\, 
    {{\cal T}^2\over 2}\, . 
    \label{5.13}
\end{eqnarray}
Due to the symmetry of the ${\bf q}_{1\perp}$- and 
${\bf q}_{2\perp}$-integrations,
one can exchange $\vec{B}^2_{1(12)} \to \vec{B}^2_{2(12)}$
in (\ref{5.13}). Also, for a direct comparison with the 
$N=2$ case (\ref{5.12}), the opacity factor ${{\cal T}^2\over 2}$ that
corresponds to the probability of two scatterings, should be dropped.

To discuss the differences between the result (\ref{5.13}) of the QCD dipole
prescription and the $N=2$ pQCD result (\ref{5.12}), we present in
Figure~\ref{fig3} the seven Feynman amplitudes contributing to the 
radiation spectrum (\ref{5.8}). For each of the diagrams, we denote
by $M_i$ the entire contribution, by $M_i^{(1)}$ the part of the
contribution whose phase factor depends only on $t_1$, by 
$M_i^{(12)}$ the part whose phase factor depends on $t_1$ and $t_2$,
etc. Based on calculations reported in Ref.~\cite{BDMPS2,GLV99},
one finds 
\begin{eqnarray}
  \vec{B}^2_1 &\propto& (M_1 + M_2^{(1)} + M_4)^2\, ,
  \label{5.14} \\
  \vec{B}^2_2 &\propto& (M_2^{(2)} + M_3 + M_6^{(2)})^2\, ,
  \label{5.15} \\
  \vec{B}^2_{2(12)} &\propto& (M_5 + M_6^{(12)} + M_7)^2\, .
  \label{5.16}
\end{eqnarray}
As can be seen from Figure~\ref{fig3}, the terms $\vec{B}^2_1$ and 
$\vec{B}^2_2$ are the Bertsch-Gunion radiation contributions off the 
first and second scattering center respectively. The term $\vec{B}^2_{2(12)}$ 
corresponds to gluon emission around $t_1$ and rescattering of the 
gluon at $t_2$. 
Since it is the gluon and not the quark which scatters off the second 
coloured potential, the term $\vec{B}^2_{2(12)}$ is enhanced by a factor 
$C_A\over C_F$ in (\ref{5.12}). These three sets of diagrams leading
to $\vec{B}^2_1$, $\vec{B}^2_2$ and $\vec{B}^2_{2(12)}$
can also be seen to form the building blocks of an effective current
in the BDMPS-approach, see e.g. Fig. 4 of 
Ref.~\cite{BDMPS2}. 
%
%
One might have expected naively that the QCD dipole prescription
maps the two Bethe-Heitler terms of (\ref{3.26}) onto the two
Bertsch-Gunion terms $\vec{B}^2_1$ and $\vec{B}^2_2$. This is only 
the case for the ${\bf k}_\perp$-integrated cross section (i.e., the
energy loss) if the kinematical boundary $\omega = x\, E_1$
is ignored. Then, the ${\bf k}_\perp$-integration allows to shift
${\bf k}_\perp \to {\bf k}_\perp - {\bf q}_{1\perp}$ which
changes $\vec{B}^2_{1(12)} \to \vec{B}^2_2$ in (\ref{5.13}). 

The $N=2$ low opacity expansion allows a quantitative check 
of QED-inspired QCD calculations. The deviation of the limiting 
case (\ref{5.13}) from the perturbative QCD result is clearly 
non-negligible: for the ${\bf k}_\perp$-integrated case, where
the $\vec{B}^2_{1(12)}$ and $\vec{B}^2_{1}$-contributions have
the same size, one can read off a correction factor
\begin{equation}
  F_{\rm corr} = 1 + {C_A\over 2\, C_F} = {17\over 8}\, ,
  \label{5.17}
\end{equation}
with which the $N=2$ dipole cross section (\ref{5.13}) has to be
multiplied to regain the pQCD radiation spectrum. [In fact, for
QCD, the ${\bf k}_\perp$-integrals are divergent in the $\Delta t \to
\infty$ limit due to the collinear singularity that must be regulated
by an in-medium screening scale.] For the ${\bf k}_\perp$-
differential spectrum or for finite $L$, corrections are clearly
more complicated, but the important message is that at least for
simple few-rescattering cases, the resulting expressions can be
compared to pQCD results and the deviations can be quantified.
The fact that the QCD dipole prescription tends to underestimate
the full pQCD radiation cross section by a significant amount 
(at least for $N=2$) may indicate that it is of use as a lower
bound for the possible radiative energy loss. 

\section{Conclusion}
\label{sec6}

In the present work, we have extended the KST-formalism~\cite{KST98} for 
quantum electrodynamics in several ways: (i) We have supplemented
it with a regularization prescription which correctly removes the
Weizs\"acker-Williams fields of the asymptotic states from the
radiation spectrum and thus warrants the correct limiting behaviour
of the formalism for a set of few-scattering scenarios. (ii) We have
calculated the corresponding angular integrated energy loss and shown
that it implies corrections to Zakharov's energy loss formula (\ref{2.47}).
(iii) We have derived a properly regularized radiation spectrum in the
dipole approximation which allows for a straightforward numerical 
calculation, as well as for further analytical studies in 
limiting cases. Most importantly, the KST-formalism describes the
angular dependence of the radiative energy loss which was shown in
section~\ref{sec4} to have a very rich non-analytical structure, 
incorporating intuitive properties of in-medium propagation as e.g.
the Moli\`ere ${\bf q}_\perp$-broadening.

In the QCD case, the situation is much more complex due to the colour
structure and the rescattering of the gluon. We have tested the QCD
dipole prescription, used in various previous 
applications~\cite{Z96,Z98,KST98,KRT98,NZ94},
against few-scattering pQCD results~\cite{GLV99}. 
The good news is that this 
prescription recovers the $N=1$ QCD Bertsch-Gunion gluon radiation 
spectrum. For multiple collisions, however, even 
the angular integrated distributions differ from the exact result 
by a sizeable factor which approaches 17/8 at large separation.
This discrepancy is due to the increase in the effective elastic
cross section because both the jet and the gluon undergo multiple
scattering. 
Clearly, the QCD problem still needs further work and the present 
formalism needs to be extended to properly take into account
non-abelian features. The introduction of the low opacity expansion
in this paper was shown to be a powerful test of any proposed
extension of QED calculations to QCD.
  
\acknowledgements
We thank Boris Kopeliovich and Andreas Sch\"afer for several discussions,
especially about the problem of regaining the Bethe-Heitler limit
from the radiation spectrum of Ref.~\cite{KST98}. After informing 
them about the importance of the regularization prescription discussed
in section~\ref{sec2b}, Boris Kopeliovich noted that Alexander Tarasov 
had come to the same conclusion on the basis of earlier calculations 
of one of us. We also thank Peter Levai and Ivan Vitev for many 
discussions about radiative energy loss calculations in the 
BDMPS-formalism and for sharing with us the unpublished result
(\ref{5.8}) whose derivation will appear in~\cite{GLV99}. Finally,
we thank U. Heinz for helpful comments regarding the manuscript. This work 
was supported by the Director, Office of Energy Research, Division of
Nuclear Physics of the Office of High Energy and Nuclear Physics of the
U.S. Department of Energy under Contract No. De-FG-02-92ER-40764.

\appendix
\section{In medium averages}
\label{appa}
In this appendix, we give details of the calculation of in medium
averages, used in the derivation of the radiation probability
(\ref{2.22}). We start from a set of $N$ static external potentials
$U_0$ at positions $({\bf r}_i,z_i)$,
\begin{equation}
  U({\bf x}) = \sum_{i=1}^N\, U_0({\bf r}-{\bf r}_i,\xi-z_i)\, .
  \label{a.1}
\end{equation}
The in medium average $\langle\cdots\rangle$ averages over the elementary
scattering centers in a volume of longitudinal extension $\Delta z$
and transverse radius $R$ according to
\begin{eqnarray}
  \langle f\rangle &\equiv& 
            \left( \prod_{i=1}^N \int_V { {\it d}z_i\, {\it d}{\bf r}_i
                               \over \pi\, R^2\, \Delta z}\right)\,
                        f({\bf r}_1,\dots,{\bf r}_N;z_1,\dots,z_N)\, ,
                        \nonumber \\
  \langle 1\rangle &=& 1\, .                         
  \label{a.2}
\end{eqnarray}     
We are interested in the combinations ${\cal F}$ of external potentials 
appearing in the squared radiation amplitude (\ref{2.18}),
 \begin{eqnarray}
   &&{\cal F}\left[{\bf r} - {\bf r}';z',z\right] =
   \Bigg\langle \exp\left\{ \sum_{i=1}^N\, 
                             W_i\bigl( {\bf r},{\bf r}';z',z\bigr)
    \right\}\Bigg\rangle\, , 
    \label{a.3} \\
   &&W_i\bigl( {\bf r},{\bf r}';z',z\bigr)
   = \nonumber \\
   && \quad 
    i\int\limits_{z}^{z'}{\it d}\xi\,  
    \left[ U\bigl({\bf r}-{\bf r}_i,\xi-z_i\bigr) 
                  - U\bigl({\bf r}'-{\bf r}_i,\xi-z_i\bigr) 
           \right]\, .
   \label{a.4}
 \end{eqnarray}
With a density $n(z_i,{\bf r}_i)$ of scattering centers
(which is $n = {N}/{\pi\, R^2\, \Delta z}$ for the homogeneous
case), the average (\ref{a.3}) can be rewritten in the form
\begin{eqnarray}
   &&\Bigg\langle \exp\left\{ \sum_{i=1}^N\, 
                              W_i \right\}\Bigg\rangle
   = \prod_{i=1}^N\, \Big\langle e^{W_i}\Big\rangle
   \nonumber \\
   && \qquad\qquad\qquad
   = \left( \frac{1}{N} \int {\it d}z_i\, {\it d}{\bf r}_i\,
              n(z_i,{\bf r}_i)\, e^{W_i} \right)^N\nonumber \\
   && \qquad\qquad\qquad
   \equiv \left( 1 + \frac{\bar{\Sigma}}{N}\right)^N
   \longrightarrow e^{\bar{\Sigma}}\nonumber \\
   && \qquad\qquad\qquad\qquad 
   \hbox{(for $N\to\infty$)}\, ,
   \label{a.5} \\
  &&\bar{\Sigma} = \int {\it d}z_i\, {\it d}{\bf r}_i\, n(z_i,{\bf r}_i)\,
              \left( 
              e^{W_i\bigl( {\bf r},{\bf r}';z',z\bigr)}-1\right)\, .
   \label{a.6}
\end{eqnarray}
The notational shorthand $\bar{\Sigma}$ introduced here does 
{\it not} grow linearly with the number of scattering centers $N$. 
This is crucial for the $N\to\infty$ limit in (\ref{a.5}) to work. 

So far, we have not specified the remaining functional dependence of
$\bar{\Sigma}$ in (\ref{a.6}). For homogeneous targets of sufficiently
large transverse size, it is clear that $\bar{\Sigma}$ can only depend 
on the relative path difference $\bbox{\rho}(\xi) = 
{\bf r}(\xi) - {\bf r}'(\xi)$. We now demonstrate for a specific
model that $\bar{\Sigma}$ takes the form
\begin{equation}
  \bar{\Sigma}[\bbox{\rho}] = - \int\limits_{z}^{z'} {\it d}\xi\,
              \Sigma\bigl(\xi,\bbox{\rho}(\xi)\bigr)\, ,
              \label{a.7}
\end{equation}
where $\Sigma\bigl(\xi,\bbox{\rho}(\xi)\bigr)$ 
$= \textstyle{1\over 2}\, n(\xi)$  $\sigma\bigl(\bbox{\rho}(\xi)\bigr)$
is defined in terms of the so-called dipole cross section 
$\sigma(\bbox{\rho})$ and the longitudinal density $n(\xi)$ of 
scattering centers, see (\ref{2.21}). 

The model under consideration is the abelian version of the Gyulassy-Wang 
model which takes for the elementary 
scattering centers in (\ref{a.1}) Yukawa potentials with
Debye mass $M$,
\begin{equation}
  U_0({\bf x}) = \frac{Z\, e^2}{4\, \pi}
                 \frac{e^{-M\, |{\bf x}|}}{|{\bf x}|}\, .
  \label{a.8}
\end{equation}
We have included in the definition of these potentials an extra
power in the coupling constant to take the coupling of the potential
to the passing electron into account. In the following calculation,
we assume that the average range $1/M$ of the potentials $U_0$ is
much smaller than the size of the medium, $z'-z \gg 1/M$. Also, we
use that density fluctuations are negligible on the scale $1/M$.
Expanding $\bar{\Sigma}$ to leading non-vanishing
order in the coupling constant $e$, we find 
\begin{eqnarray}
   \sigma\bigl(\bbox{\rho}(\xi)\bigr)
          &=& 2\, \int\limits_{-\infty}^{\infty} {\it d}\xi\,
              \left[ 
              \langle\!\langle U_0(\bbox{0},0)\, U_0(\bbox{0},\xi')
              \rangle\!\rangle \right. 
              \nonumber \\
           && \qquad \qquad - \left. \langle\!\langle U_0(\bbox{0},0)\, 
               U_0(\bbox{\rho}(\xi),\xi') \rangle\!\rangle 
              \right] \, ,
  \label{a.9}
\end{eqnarray}
where
\begin{equation}
  \langle\!\langle U_0(0)\, U_0(\bbox{x}) \rangle\!\rangle 
  \equiv \int {\it d}\bbox{\bar x }\, U_0(\bbox{\bar x})\,
                U_0(\bbox{\bar x}-\bbox{x})\, .
  \label{a.10}
\end{equation}
The integrals in (\ref{a.9}) can be done analytically, leading to
Zakharov's result~\cite{Z87}
\begin{equation}
  \sigma(\bbox{\rho}) =  
         \frac{Z^2\, \alpha^2_{\rm em}}{M^2}\, 8\, \pi\,
         \bigl( 1 - M\,\rho\, K_1(M\, \rho)\bigr) \, .
  \label{a.11}
\end{equation}
In our study of the ${\bf k}_\perp$-dependence of the radiation
probability, we shall encounter the Fourier transform of 
$\bar{\Sigma}[\bbox{\rho}]$ with respect to the transverse momentum
transfer ${\bf q}_\perp$ from the medium to the electron, 
\begin{eqnarray}
  \tilde{\Sigma}({\bf q}_\perp) 
  &=&\frac{1}{2}\, \int \frac{ {\it d}\bbox{\rho}}{(2\,\pi)^2}\,
      \sigma(\bbox{\rho})\, 
      \exp\left\{i\, {\bf q}_\perp\cdot \bbox{\rho} \right\}
      \nonumber \\
  && = - \frac{4\,Z^2\, \alpha_{\rm em}^2}{(M^2+{\bf q}_\perp^2)^2}
     + \frac{4\,\pi\,Z^2\, \alpha_{\rm em}^2}{M^2}\, 
       \delta^{(2)}({\bf q}_\perp)\, .
  \label{a.12}
\end{eqnarray}
For non-vanishing momentum transfer, the first term of (\ref{a.12}) 
is the Mott cross-section for electron scattering off the single 
external potential (\ref{a.1}) in the relativistic limit. The size
of the prefactor of the $\delta^{(2)}({\bf q}_\perp)$ contribution
is such that the ${\bf q}_\perp$-integral over (\ref{a.12}) vanishes.
This warrants that the dipole cross section $\sigma(\bbox{\rho})$ 
vanishes at vanishing dipole size $\rho = 0$.
One can trace back the origin of this second term:
in the expansion of (\ref{a.4}) to second order in the coupling
constant, it arises from contributions which are second order in 
$U(\bbox{r}-\bbox{r}_i)$ but zeroth order in $U(\bbox{r}'-\bbox{r}_i)$,
or vice versa. This amounts to multiplying the second order term
of $M^{(2)}_{fi}$ with the zeroth order term ${M^{(0)}_{fi}}^*$
of the complex conjugated amplitude. The contribution $M^{(2)}_{fi}$
$\times\, {M^{(0)}_{fi}}^*$ to the radiation cross section vanishes
however due to energy momentum conservation. Calculating such
a contribution from the Furry approximation (\ref{2.5}), 
(\ref{2.6}), this energy momentum conservation appears in the
form of a $\delta$-function $\delta^{(2)}({\bf q}_\perp)$, 
resulting in an additional contribution to the Mott cross section.
For a comparison with experimental data, one has to assume a
transverse momentum transfer $\bbox{q}_\perp$ which is always finite
but can become arbitrarily small. Then the second term in (\ref{a.12})
drops out in all physical quantities.

\section{Eikonal expressions from path integrals}
\label{appb}
In this appendix, we give details of the solutions of the path
integrals, used for the derivation of the radiation probability
$\langle|M_{fi}|^2\rangle$ in (\ref{2.22}). We start from the 
observation that with the help of the averaging (\ref{2.20})
\begin{eqnarray}
 && {\cal S}(z',{\bf r}(z'),{\bf r'}(z');z,{\bf r}(z),{\bf r'}(z)|p)
   \nonumber \\
 &&\equiv \langle G(z',{\bf r}(z');z,{\bf r}(z)|p)\,\,
                  G^*(z',{\bf r'}(z');z,{\bf r'}(z)|p)\, \rangle
 \nonumber \\
 && = \int {\cal D}\bbox{\bar\rho}\, {\cal D}\bbox{\hat\rho}\,
 \exp\left\{ \frac{ip}{2} \int_z^{z'} 
  \bbox{ \dot {\hat\rho} } \cdot
  \bbox{ \dot {\bar\rho} } 
 \right\}\, {\cal F}\left[\bbox{\hat\rho};z',z\right]\, , 
 \label{b.1} \\
 && \bbox{\hat\rho}(z) = \bbox{r}(z) - \bbox{r}'(z)\, ,
    \qquad  \bbox{\bar\rho}(z) = \bbox{r}(z) + \bbox{r}'(z)\, .
 \label{b.2}
\end{eqnarray}
%
%
%
%
Here, the dot denotes a derivative with respect to the longitudinal
direction and the integral $\int {\cal D}\bbox{\bar\rho}$ is trivial.
It results in a $\delta$-function which constrains all possible
paths $\bbox{\hat\rho}(\xi)$ by the mid-point rule to a  straight line 
\begin{equation}
  \bbox{\hat\rho}_s(\xi) = \bbox{\hat\rho}(z')\frac{\xi-z}{z'-z}
                         + \bbox{\hat\rho}(z)\frac{z'-\xi}{z'-z}\, .
  \label{b.3}
\end{equation}
The final result is
\begin{eqnarray}
 &&{\cal S}(z',{\bf r}(z'),{\bf r'}(z');z,{\bf r}(z),{\bf r'}(z)|p)
  \nonumber \\
 && = - \left( {p\over 2\pi i (z'-z)}\right)^2
    \exp\left\{ - \int_z^{z'} 
                \Sigma\bigl(\xi,\bbox{\hat\rho}_s(\xi)\bigr) {\it d}\xi
        \right\} 
 \nonumber \\
 && \times \exp\left\{ { {ip}\,    
 \left[ \bigl(\bbox{r}(z')-\bbox{r}(z)\bigr)^2
        - \bigl(\bbox{r}'(z')-\bbox{r}'(z)\bigr)^2 \right]
 \over {2\, (z'-z)} } 
 \right\}\, ,
 \label{b.4} 
\end{eqnarray}
We note that this `eikonal path' $\bbox{\hat\rho}_s$ does not 
result from an eikonal approximation. It is rather obtained from a 
calculation on the cross section level, once the ingoing and outgoing
states $\Psi_F^{\pm}$ of (\ref{2.5}), (\ref{2.6}) are adopted as
starting point: the Furry approximation is a systematic expansion in
$1/E$ and thus closely related to the eikonal approximation.
%
\begin{figure}[h]\epsfxsize=8.5cm 
\centerline{\epsfbox{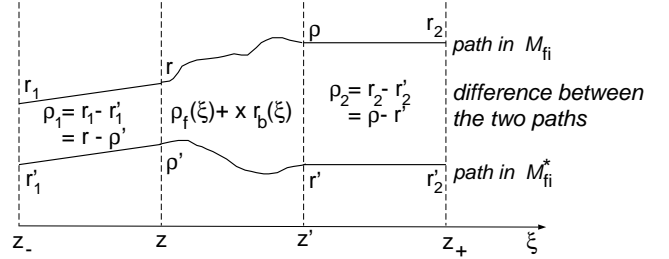}}
\caption{Diagrammatic representation of the path integrals appearing
in the radiation probability (\protect\ref{2.22}). According to 
(\protect\ref{b.1}), the path differences ${\rho}_1$ and 
${\rho}_2$ at early times $\xi < z$ and late times $\xi > z'$
are fixed to straight lines. Between $z$ and $z'$, the path 
difference can change due to interactions with the medium, but
the boundary conditions are fixed.  
}\label{fig1}
\end{figure}
%
The explicit expression (\ref{b.4}) allows to simplify
the radiation probability (\ref{2.18}) by using
\begin{eqnarray}
  &&\int d{\bf r}_2\, d{\bf r}'_2\, 
  e^{-i{\bf p}_{2\perp}\cdot ({\bf r}_2-{\bf r}'_2)}
  S(z_+,\bbox{r}_2,\bbox{r}'_2; z',\bbox{\rho},\bbox{r}'|p_2)
  \nonumber \\
  && = - \exp\left\{-i\bbox{p}_{2\perp}\cdot \bbox{\rho}_2 -
       \int_{z'}^{z_+} \Sigma(\xi,\bbox{\rho}_2)\, {\it d}\xi\right\}\, ,
  \label{b.5} \\
  &&\int d{\bf r}_1\, d{\bf r}'_1\, 
  e^{i{\bf p}_{1\perp}\cdot ({\bf r}_1-{\bf r}'_1)}
  S(z,\bbox{r},\bbox{\rho}';z_-,\bbox{r}_1,\bbox{r}'_1|p_1)
  \nonumber \\
   && = - \exp\left\{i\bbox{p}_{1\perp}\cdot \bbox{\rho}_1 -
       \int_{z_-}^{z} \Sigma(\xi,\bbox{\rho}_1)\, {\it d}\xi\right\}\, ,
  \label{b.6} 
\end{eqnarray}
where we have introduced the coordinates
\begin{eqnarray}
  \bbox{\rho}_1 &=& \bbox{r} - \bbox{\rho}'\, ,
  \qquad
  \bbox{\bar\rho}_1 = \bbox{r} + \bbox{\rho}'\, ,
  \nonumber \\
  \bbox{\rho}_2 &=& \bbox{\rho} - \bbox{r}'\, ,
  \qquad
  \bbox{\bar\rho}_2 = \bbox{\rho} + \bbox{r}'\, .
  \label{b.7}
\end{eqnarray}
The integrals (\ref{b.5}), (\ref{b.6}) depend only on the relative
distances $\bbox{\rho}_1$, $\bbox{\rho}_2$. Also, 
due to the boundary conditions, the path $\bbox{\hat\rho}_s(\xi)$ 
turns out to be $\xi$-independent in (\ref{b.5}) and
(\ref{b.6}).

Equation (\ref{b.4}) provides an explicit expression for the in-medium
average of pairs of Green's functions as long as both Green's functions
are for particles with the same momentum $p$. To calculate the 
radiation probability (\ref{2.17}), we have to extend this calculation
to the case of two different momenta. To this end, we use the identity
\begin{equation}
  1 = {\cal N}(\bbox{r},\bbox{\rho})\, \int {\cal D}\bbox{r}_e\, 
  \exp\left\{ \frac{ip}{2} x \int_z^{z'} 
  \bbox{\dot r}_e^2(\xi) {\it d}\xi \right\}\, ,
  \label{b.8}
\end{equation}
where the norm ${\cal N}(\bbox{r},\bbox{\rho})$ is given in terms of 
the boundary conditions for the path integral over $\bbox{r}_e$,
\begin{eqnarray}
  {\cal N}(\bbox{r},\bbox{\rho}) &=& \frac{2\pi i\, (z'-z)}{p\, x}\, 
  \exp\left\{ -\frac{i\, p\, x}{2\, (z'-z)} 
              \left(\bbox{r} - \bbox{\rho}\right)^2 \right\}\, ,
  \label{b.9} \\
  && \bbox{r}_e(z) \equiv \bbox{r}(z) = \bbox{r}\, , \qquad 
  \bbox{r}_e(z') \equiv \bbox{r}(z') = \bbox{\rho}\, .
  \label{b.10}
\end{eqnarray}
This allows us to write 
\begin{eqnarray}
 &&\langle G(z',\bbox{\rho};z,\bbox{r}|p(1-x))\,\,
                  G^*(z',\bbox{r}';z,\bbox{\rho}'|p)\, \rangle
  \nonumber \\
 && = {\cal N}(\bbox{r},\bbox{\rho})\, 
      \int {\cal D}\bbox{r}\, {\cal D}\bbox{r}'\, {\cal D}\bbox{r}_e\,
      {\cal F}(\bbox{r}-\bbox{r}')
 \nonumber \\
 && \qquad \times \exp\left\{ \frac{ip}{2} \int_z^{z'}    
 \left[ (1-x)\, \bbox{\dot r}^2 + x\, \bbox{\dot r}_e^2\, 
        - \bbox{\dot r}'^2 \right] 
 \right\}\, ,
 \label{b.11} 
\end{eqnarray}
where $\bbox{r}'(z') = \bbox{r}'$ and $\bbox{r}'(z) = \bbox{\rho}'$.
Introducing the new coordinates
\begin{eqnarray}
  \bbox{r}_a(\xi) &=& (1-x)\, \bbox{r}(\xi) + x\, \bbox{r}_e(\xi)\, ,
  \label{b.12}\\
  \bbox{r}_b(\xi) &=& \bbox{r}(\xi) - \bbox{r}_e(\xi)\, ,
  \label{b.13}
\end{eqnarray}
the exponent of the path integral (\ref{b.10}) reads
\begin{equation}
  (1-x)\, \bbox{\dot r}^2 + x\, \bbox{\dot r}_e^2\, 
        - \bbox{\dot r}'^2 
  =  \bbox{\dot r}_a^2 - \bbox{\dot r}'^2\, 
        + (1-x)x\bbox{\dot r}_b^2\, ,
  \label{b.14}
\end{equation}
and ${\cal F}$ appears with the argument
${\cal F}(\bbox{r}_a - \bbox{r}' + x\, \bbox{r}_b)$. In
complete analogy to the calculation of (\ref{b.4}), this
renders the path integral over $(\bbox{r}_a + \bbox{r}')$
trivial and reduces the support of $(\bbox{r}_a - \bbox{r}')$
to a straight line. With the choice of boundary conditions (\ref{b.9}),
we find 
\begin{eqnarray}
&&\langle G(z',\bbox{\rho};z,\bbox{r}|p_2)\,\,
                  G^*(z',\bbox{r}';z,\bbox{\rho}'|p_1)\, \rangle
   = \frac{-p}{2\pi\, i\, (z'-z)} \frac{1}{x}
 \nonumber \\
 &&
    \times \exp\left\{ \frac{ip}{2\, (z'-z)} 
      \left[ (1-x) \bigl(\bbox{\rho} - \bbox{r}\bigr)^2
             -  \bigl(\bbox{r}' - \bbox{\rho}'\bigr)^2 \right]
             \right\}
 \nonumber \\
 && \times \int {\cal D}\bbox{r}_b
    \exp\left\{ \frac{i\mu}{2}\int_z^{z'}\bbox{\dot r}_b^2
                - \int_z^{z'} \Sigma\bigl(\xi,\bbox{\hat \rho}_f 
                + x\bbox{r}_b\bigr)
        \right\}\, ,
 \label{b.15}
\end{eqnarray}
where $\mu = p\, (1-x)\, x$ and 
\begin{eqnarray}
    \bbox{\hat\rho}_f(\xi) = \bbox{\rho}_2
                             \frac{\xi-z}{z'-z}
                           + 
                             \bbox{\rho}_1
                             \frac{z'-\xi}{z'-z}\, .
  \label{b.16}
\end{eqnarray}

We can now simplify the radiation probability $\langle|M_{fi}|^2\rangle$
of (\ref{2.18}) with the help of (\ref{b.5}), (\ref{b.6}) and (\ref{b.15}).
To this aim, we observe first that with a partial integration, the 
derivative $\partial /\partial\bbox{r}$ $\partial /\partial\bbox{r}'$ 
of the interaction vertex (\ref{2.17}) changes into
$-\partial /\partial\bbox{\rho}_1$ $\partial /\partial\bbox{\rho}_2$.
Then we change to the integration variables
\begin{eqnarray}
  \bbox{\tau} &=& \bbox{\bar\rho}_1 - \bbox{\bar\rho}_2\, ,
  \qquad 
  \bbox{\bar\tau} = \bbox{\bar\rho}_1 + \bbox{\bar\rho}_2\, ,
  \nonumber \\
  \bbox{b} &=& \frac{1}{4}\bbox{\bar\tau} 
            = \frac{1}{4} \left( \bbox{r} + \bbox{\rho}' + 
                                 \bbox{\rho} + \bbox{r}' \right)\, .
  \label{b.17}
\end{eqnarray}
The only $\bbox{\tau}$-dependence of the integrand of 
$\langle|M_{fi}|^2\rangle$ is in the exponential of the
second line of (\ref{b.15}). Rewriting this in the
new integration variables (\ref{b.7}) and (\ref{b.17}),
we find
\begin{eqnarray}
   && \int {\it d}\bbox{\tau} \exp\left\{ { {ip}
           \left[ (\bbox{\rho}_1-\bbox{\rho}_2)\cdot\bbox{\tau}
                 -\frac{x}{4} \left( (\bbox{\rho}_1-\bbox{\rho}_2) 
                                     + \bbox{\tau}\right)^2
           \right] \over {2(z'-z)} }
           \right\} \nonumber\\
   && = \frac{- 8\pi i\, (z'-z)}{p\, x}
       \exp\left\{ \frac{ip\, (1-x)}{2(z'-z)x} 
             \left(\bbox{\rho}_1-\bbox{\rho}_2\right)^2 \right\}\, . 
   \label{b.18}
\end{eqnarray}
The radiation probability reads now
\begin{eqnarray}
 &&\langle|M_{fi}|^2\rangle = \frac{8}{x^2} 
 {\rm Re}\, 
 \int {\it d}\bbox{\rho}_1\, {\it d}\bbox{\rho}_2\, {\it d}\bbox{b}
      \int dz\, \int_z^\infty dz'\, 
      \nonumber \\
  &&\quad \times 
      \exp\left\{i\bar{q}(z'-z) + i\bbox{p}_{1\perp}\cdot \bbox{\rho}_1
      -i\bbox{p}_{2\perp}\cdot \bbox{\rho}_2
      \right\}
      \nonumber \\
  &&\quad \times \exp\left\{-\int_{z_-}^{z} 
                 \Sigma(\xi,\bbox{\rho}_1)\, {\it d}\xi
                -\int_{z'}^{z_+} 
                 \Sigma(\xi,\bbox{\rho}_2)\, {\it d}\xi 
          \right\} 
       \nonumber \\
  &&\quad \times      
  \widehat{\Gamma}_{-{\rho}_1}\,
  \widehat{\Gamma}_{-{\rho}_2}^*\,
       \exp\left\{ \frac{ip\, (1-x)}{2(z'-z)x} 
             \left(\bbox{\rho}_1-\bbox{\rho}_2\right)^2 \right\}
     \nonumber \\
  &&\quad \times \int {\cal D}\bbox{r}_b\,
         \exp\left\{ \frac{i\, p\, (1-x)\, x}{2}
                  \int_z^{z'}\bbox{\dot r}_b^2(\xi)\, {\it d}\xi
                  \right\}
      \nonumber \\
  && \qquad \quad \times
               \exp\left\{  
                - \int_z^{z'} \Sigma\bigl(\xi,\bbox{\hat \rho}_f 
                + x\bbox{r}_b\bigr)\, {\it d}\xi
        \right\}\, .
  \label{b.19}
\end{eqnarray}
This is proportional to a transverse area ${\it d}\bbox{b}$ which 
we devide out in what follows: the corresponding radiation spectrum
is given per unit transverse area. The above expression
can be simplified further if one observes that
\begin{equation}
  \int_z^{z'} \left( \bbox{\dot {\hat\rho}}_f + \bbox{\dot r}_b\right)^2 
  {\it d}\xi = 
  \frac{\left(\bbox{\rho}_1-\bbox{\rho}_2\right)^2}{(z'-z)}
  + \int_z^{z'} \bbox{\dot r}_b^2 {\it d}\xi\, .
  \label{b.20}
\end{equation}
After a transformation $\bbox{\rho}_1 \to x\, \bbox{\rho}_1$, 
and $\bbox{\rho}_2 \to x\, \bbox{\rho}_2$, this allows
to combine in (\ref{b.19}) the Gaussian exponent and the path
integral into a new path integral over
\begin{equation}
  \bbox{r}_c(\xi) = \bbox{\hat\rho}_f(\xi) + \bbox{r}_c(\xi)\, ,
  \label{b.21}
\end{equation}
with boundary conditions
\begin{eqnarray}
  \bbox{r}_c(z) &=& \bbox{\hat\rho}_f(z) = \bbox{\rho}_1\, ,
  \nonumber \\
  \bbox{r}_c(z') &=& \bbox{\hat\rho}_f(z') = \bbox{\rho}_2\, .
  \label{b.22}
\end{eqnarray}
The resulting expression is given in (\ref{2.22}) and differs
from the result of Ref.~\cite{KST98} by the regularization
$\exp\left\{ -\epsilon\, |z|\right\}$ of the integrand only. 
The extension of the formalism to spin- and helicity dependent
quantities is obtained, by inserting for $\widehat{\Gamma}_{-{\rho}_1}$
and $\widehat{\Gamma}_{-{\rho}_2}^*$ in (\ref{b.19}) the spin- and
helicity dependent vertex functions (\ref{2.15}) and (\ref{2.16}),
rather than the average (\ref{2.17}). 

\section{Multiple scatterings: N=3}
\label{appc}
For the case of $N=3$ scatterings,
the combined momentum transfer ${\bf q}_\perp$ from the medium is 
desribed by the folding of three elastic Mott cross sections 
$\tilde{\Sigma}$,
\begin{eqnarray}
  &&\int d{\cal V}^3(\bbox{q}_\perp) \equiv  
  - \int d{\bf q}_{1\perp}\, d{\bf q}_{2\perp}\, d{\bf q}_{3\perp}\, 
  \tilde{\Sigma}({\bf q}_{1\perp})\, \tilde{\Sigma}({\bf q}_{2\perp})\,
  \tilde{\Sigma}({\bf q}_{3\perp})
  \nonumber \\
  && \qquad \times {(2\pi)^2\over x^2} \delta^{(2)}\left({\bf q}_\perp - 
     {\bf q}_{1\perp} - {\bf q}_{2\perp} - {\bf q}_{3\perp}\right)\, .
  \label{3.27}
\end{eqnarray}
With the shorthands 
\begin{eqnarray}
  {\bf u}_{m1} &=& x\,\left({\bf p}_{1\perp} + {\bf q}_{1\perp}\right)\, ,
  \nonumber \\
  {\bf u}_{m2} &=& x\,\left({\bf p}_{1\perp} 
  + {\bf q}_{1\perp} + {\bf q}_{2\perp}\right)\, ,
  \nonumber \\
  Q_{m1} &=& { {\bf u}_{m1}^2\over 2\mu}\, ,\qquad
  Q_{m2} = { {\bf u}_{m2}^2\over 2\mu}\, ,
  \label{3.28} 
\end{eqnarray}
the radiation spectrum (\ref{2.22}) expanded to third order reads
\begin{eqnarray}
 && \frac{d^5\sigma}{d({\rm ln}x)\,d{\bf p}_\perp\,d{\bf k}_\perp}
 \Bigg\vert_{O(\alpha_{\rm em}^7\, {\cal T}^3)} 
 = C_{\rm pre}\, g_{\rm nf}\, \int d{\cal V}^3(\bbox{q}_\perp)
 \nonumber \\
 && \qquad \times \left[ {1\over 3!}\, {\bf u}_2^2 {\cal Z}_1^{(3)}  
            + {1\over 2}\,{\bf u}_{m2}^2 {\cal Z}_2^{(3)}
            + {1\over 2}\,{\bf u}_{m1}^2 {\cal Z}_3^{(3)}
            + {1\over 3!}\, {\bf u}_1^2 {\cal Z}_4^{(3)} \right.
          \nonumber \\
 && \qquad\qquad 
            + {1\over 2}\, {\bf u}_2\cdot{\bf u}_{m2} {\cal Z}_5^{(3)}
            + {1\over 2}\, {\bf u}_1\cdot{\bf u}_{m1} {\cal Z}_6^{(3)}
          \nonumber \\
 && \qquad\qquad 
            + {\bf u}_{m1}\cdot{\bf u}_{m2} {\cal Z}_7^{(3)}
            + {\bf u}_{2}\cdot{\bf u}_{m1} {\cal Z}_8^{(3)} 
          \nonumber \\
 && \qquad\qquad \left. 
            + {\bf u}_{m2}\cdot{\bf u}_{1} {\cal Z}_9^{(3)}
            + {\bf u}_{1}\cdot{\bf u}_{2} {\cal Z}_{10}^{(3)}
            \right]\, n^3_0\, .
  \label{3.29}
\end{eqnarray}
The variables ${\cal Z}^{(3)}_i$ stand again for the longitudinal
integrals over phase factors. For a homogeneous medium of density
$n_0$ and length $L$ they read
\begin{eqnarray}
  {\cal Z}_1^{(3)} &=& {L^3\over 2\, Q_2^2}\, ,
  \label{3.30} \\
  {\cal Z}_2^{(3)} &=& {{6\,\sin(L\, Q_{m2}) - 6\, L\, Q_{m2} + 
                        L^3\, Q_{m2}^3}\over 3\, Q_{m2}^5}\, ,
  \label{3.31} \\
  {\cal Z}_3^{(3)} &=& {{6\,\sin(L\, Q_{m1}) - 6\, L\, Q_{m1} + 
                        L^3\, Q_{m1}^3}\over 3\, Q_{m1}^5}\, ,
  \label{3.32} \\
  {\cal Z}_4^{(3)} &=& {L^3\over 2\, Q_1^2}\, ,
  \label{3.33} \\
  {\cal Z}_5^{(3)} &=& -{{6\,\sin(L\, Q_{m2}) - 6\, L\, Q_{m2} + 
                        L^3\, Q_{m2}^3}\over 3\, Q_2\, Q_{m2}^4}\, ,
  \label{3.34} \\
  {\cal Z}_6^{(3)} &=& -{{6\,\sin(L\, Q_{m1}) - 6\, L\, Q_{m1} + 
                        L^3\, Q_{m1}^3}\over 3\, Q_1\, Q_{m1}^4}\, ,
  \label{3.35} \\
  {\cal Z}_7^{(3)} &=&  - {L^3\over 6\, Q_1\, Q_2}\, ,
                       + { {\left(\sin(L\, Q_{m1}) - L\, Q_{m1}\right)}
                         \over Q_{m1}^4\, \left(Q_{m1}-Q_{m2}\right)}
  \nonumber \\
                   &&  - { {\left(\sin(L\, Q_{m2}) - L\, Q_{m2}\right)}
                         \over \left(Q_{m1}-Q_{m2}\right)\, Q_{m2}^4}\, ,
  \label{3.36} \\
  {\cal Z}_8^{(3)} &=& { {\left(\sin(L\, Q_{m2}) - L\, Q_{m2}\right)}
                         \over Q_2\, \left(Q_{m1}-Q_{m2}\right)
                               \, Q_{m2}^3}
  \nonumber \\
                   &&  - { {\left(\sin(L\, Q_{m1}) - L\, Q_{m1}\right)}
                        \over Q_2\, Q_{m1}^3\, \left(Q_{m1}-Q_{m2}\right)}\, ,
  \label{3.37} \\
  {\cal Z}_9^{(3)} &=&{ {\left(\sin(L\, Q_{m2}) - L\, Q_{m2}\right)}
                        \over Q_1\, \left(Q_{m1}-Q_{m2}\right)
                               \, Q_{m2}^3}
  \nonumber \\ 
                   && - { {\left(\sin(L\, Q_{m1}) - L\, Q_{m1}\right)}
                       \over Q_1\, Q_{m1}^3\, \left(Q_{m1}-Q_{m2}\right)}\, ,
  \label{3.38} \\
  {\cal Z}_{10}^{(3)} &=& { {\left(\sin(L\, Q_{m1}) - L\, Q_{m1}\right)}
                          \over Q_1\, Q_2\, Q_{m1}^2\, 
                               \left(Q_{m1}-Q_{m2}\right)}
  \nonumber \\
                   && - { {\left(\sin(L\, Q_{m2}) - L\, Q_{m2}\right)}
                         \over Q_1\, Q_2\, \left(Q_{m1}-Q_{m2}\right)
                               \, Q_{m2}^2}\, .
  \label{3.39}
\end{eqnarray}
From this one finds again the simple limiting cases. We confirm the
factorization limit
\begin{eqnarray}
 && \lim_{L\to 0} 
 \frac{d^5\sigma}{d({\rm ln}x)\,d{\bf p}_\perp\,d{\bf k}_\perp}
 \Bigg\vert_{O(\alpha_{\rm em}^7\, {\cal T}^3)}^{{\cal T}\, {\rm fixed}} 
  \nonumber \\
 && \quad = C_{\rm pre}\, g_{\rm nf}\, 
 { {\cal T}^3\over 12}
 \left( { {\bf u}_1\over Q_1} - {{\bf u}_2\over Q_2} \right)^2\, 
 \int d{\cal V}^3(\bbox{q}_\perp)\, ,
  \label{3.40}
\end{eqnarray}
and the Bethe-Heitler limit
\begin{eqnarray}
 && \lim_{L\to \infty} 
 \frac{d^5\sigma}{d({\rm ln}x)\,d{\bf p}_\perp\,d{\bf k}_\perp}
 \Bigg\vert_{O(\alpha_{\rm em}^7\, {\cal T}^3)}^{{\cal T}\, {\rm fixed}} 
 = C_{\rm pre}\, g_{\rm nf}\, \int d{\cal V}^3(\bbox{q}_\perp)
 \nonumber \\
 && \qquad \times
 { {\cal T}^3\over 12} \left[
 \left( { {\bf u}_1\over Q_1} - {{\bf u}_{m1}\over Q_{m1}} \right)^2
 + \left( { {\bf u}_{m1}\over Q_{m1}} 
           - {{\bf u}_{m2}\over Q_{m2}} \right)^2 \right.
 \nonumber \\
 && \qquad \left. + 
 \left( { {\bf u}_{m2}\over Q_{m2}} - {{\bf u}_2\over Q_2} \right)^2
 \right]\, .
  \label{3.41}
\end{eqnarray}
Also, we have checked in complete analogy to the case of $N=2$ scatterings 
that the lowest order correction to the factorization limit (\ref{3.40}) 
is proportional to $L^2$ times ${\cal T}^3$. The ${\bf q}_\perp$-integral
of this correction diverges logarithmically. This further corroborates
the conclusions drawn in subsection~\ref{sec3b}.

%

\end{document}